\documentclass{article}
\usepackage[dvipdfmx]{graphicx}
\usepackage{amsmath}
\usepackage{amssymb}
\usepackage{authblk}
\usepackage{mathrsfs}
\usepackage{slashed}
\usepackage{color}
\usepackage{cite}
\usepackage{here}
\usepackage{subcaption}

\makeatletter
 
 \@addtoreset{equation}{section}
 \makeatother

\newcommand{\bea}{\begin{equnarray}}
\newcommand{\eea}{\end{eqnarray}}
\newcommand{\beann}{\begin{equnarray*}}

\newcommand{\eeann}{\end{eqnarray*}}
\newcommand{\nn}{\nonumber}
\newcommand{\ba}{\begin{array}}
\newcommand{\ea}{\end{array}}

\newcommand{\bs}{\boldsymbol}

\newcommand{\bse}{{\boldsymbol{e}}}
\newcommand{\bsone}{\bs{1}}

\newcommand{\lrpar}[1]{\left({#1}\right)}
\newcommand{\sgn}{{\rm sgn}\,}
\newcommand{\N}{\mathbb N}
\newcommand{\Z}{\mathbb Z}

\newcommand{\tC}{{\tilde{C}}}

\newcommand{\cB}{{\cal B}}

\newcommand{\MM}{\frac{1}{N n_V}\Bigl\langle\sum_{v\in V} {\rm Tr} \Phi_v^2  \Bigr\rangle}

\DeclareMathOperator{\Tr}{Tr}

\title{Graph Zeta Functions\\and\\Wilson Loops in Kazakov-Migdal Model}
\author[1]{So Matsuura\thanks{s.matsu@keio.jp}}
\author[2]{Kazutoshi Ohta\thanks{kohta@law.meijigakuin.ac.jp}}
\affil[1]{\it Hiyoshi Departments of Physics,
and Research and Education Center for Natural Sciences,
Keio University, 4-1-1 Hiyoshi, Yokohama, Kanagawa 223-8521, Japan}
\affil[2]{\it Institute of Physics, Meiji Gakuin University, Yokohama, Kanagawa 244-8539, Japan}

\date{}

\begin{document}
\maketitle

\vspace*{2cm}

\begin{center}
{\bf Abstract}
\end{center}

In this paper, we consider an extended Kazakov-Migdal model defined on an arbitrary graph.
The partition function of the model, which is expressed as the summation of all Wilson loops on the graph, turns out to be represented by the Bartholdi zeta function weighted by unitary matrices on the edges of the graph. 
The partition function on the cycle graph at finite $N$ is expressed by the generating function of the generalized Catalan numbers.
The partition function on an arbitrary graph can be exactly evaluated at large $N$ which is expressed as an infinite product of a kind of deformed Ihara zeta function. 
The non-zero area Wilson loops do not contribute to the leading part of the $1/N$-expansion of the free energy but to the next leading. 
The semi-circle distribution of the eigenvalues of the scalar fields is still an exact solution of the model at large $N$ on an arbitrary regular graph, 
but it reflects only zero-area Wilson loops.


\newpage

\section{Introduction}
\label{sec:Introduction}


Wilson loops \cite{PhysRevD.10.2445} are key objects to understand the non-perturbative dynamics of non-Abelian gauge theories. 
Since the Wilson loop (and the Polyakov loop) is related to the potential between the (infinitely heavy) quark and anti-quark, it is an important observable for approaching the quark confinement problem, which is one of the main subjects of the non-Abelian gauge theory.

The Wilson loops are the most general gauge invariant operators in non-Abelian gauge theory. 
In fact, the gauge theory can even be reformulated in terms of the Wilson loops \cite{makeenko1979exact,makeenko1981quantum}. 
It is therefore not a coincidence that the simplest action of the lattice gauge theory is given by the plaquette variable, that is, the smallest Wilson loop on the lattice.
When the Wilson loop on the lattice is expanded in terms of the lattice spacing, each term is expressed as a product of the field strengths of the gauge fields. 
While the leading term of the expansion of the plaquette action gives the 
gauge kinetic term  $\Tr F^2$ in the continuous theory,
the expansion of larger Wilson loops begins with higher-derivative operators. 
Thus, if we construct a general gauge invariant action on the lattice by adding up several Wilson loops, the irrelevant higher-derivative terms do not contribute in the continuum limit in general and it is considered to describe the non-Abelian gauge theories on the continuous space-time universally. 

The so-called Kazakov-Migdal (KM) model \cite{kazakov1993induced} is a lattice gauge theory defined on the $D$-dimensional lattice with the action,
\begin{equation}
  S_{\rm KM}= N \sum_x \Tr\left\{\frac{m^2}{2} \Phi^2(x) - \sum_{\mu=1}^D
  \Phi(x)U_\mu(x)\Phi(x+\mu) U_\mu^\dagger(x)
  \right\}\,,
  \label{eq:SKM}
\end{equation}
where $U_\mu(x)$ is a unitary variable living on the link extending from the site (vertex) $x$ to the direction of $\mu$ and
$\Phi(x)$ is a scalar field in the adjoint representation of $SU(N)$ living on $x$.
Remarkably, the effective action obtained by integrating out the scalar fields of this model is the sum of
all possible Wilson loops. 
Furthermore, this model can be solved exactly at large $N$ thanks to its simple structure \cite{migdal1993exact,gross1992some}.
Hence, 
this model first attracted attention as a theory that can be solved at large $N$ and is expected to induce QCD (more precisely, quantum ``gluodynamics'' without quarks)
\cite{migdal19931, gocksch1992phase, khokhlachev1992problem, caselle1992exact, aoki1993spectrum, aoki1993study, makeenko1993exact}.

However, it has become clear that the KM model does {\em not} actually induce QCD \cite{kogan1992induced,kogan1993area,kogan1993continuum}.
The main difficulty is that this model has an extra $\Z_N$ gauge symmetry%
\footnote{
For the gauge group is $U(N)$, it becomes an $U(1)$ center of the gauge symmetry. 
};
$U_\mu(x)\to \omega_\mu(x) U_\mu(x)$ so that $\omega_\mu(x)^N=1$,
which is not provided by Wilson's lattice gauge theory \cite{kogan1992induced}.
Due to this symmetry, 
a natural operator with finite expected value is not a single Wilson loop $\langle \Tr W_C(U) \rangle$,
which trivially vanishes, 
but the square of its absolute value  $\langle |\Tr W_C(U)|^2 \rangle$. 
This is not the situation realized by the usual QCD.
Despite various efforts to circumvent this difficulty, there was no sign of QCD being realized in the continuous limit of modified KM models \cite{khokhlachev1992adjoint, migdal1993mixed, migdal1993bose,cline1993induced,balakrishna1994difficulties}.

Even if the KM does not induce QCD, however, this model is still an interesting matrix model in its own right 
\cite{boulatov1993infinite,Arefeva:1993ik,caselle1993kazakov,caselle1994two,makeenko1993some,paniak1995kazakov}.
In particular, it is still true that the KM model counts all possible Wilson loops on the lattice in the form of the square of the absolute value. 
In our previous paper \cite{matsuura2022kazakov},
we proposed a generalized KM model which is defined on an arbitrary graph, 
which is called the graph KM (gKM) model. 
We showed that, by tuning the parameters of the model appropriately, 
the partition function is expressed as the integral of the Ihara zeta function \cite{Ihara:original,MR607504,sunada1986functions} weighted by unitary matrices on the edges of the graph.
The Ihara zeta function is the simplest of the graph zeta functions, 
which counts all reduced cycles on the graph.
Correspondingly, we can explicitly write down the effective action of this parameter-tuned model as a theory that adds up all non-zero area Wilson loops on a given graph.
In particular, we can evaluate the partition function exactly in the large $N$ limit thanks to the large $N$ factorization of the Wilson loops and properties of the Ihara zeta function, which becomes an infinite product of the (normal) Ihara zeta function in general. 
This result suggests that the graph zeta function will be useful for the analysis of lattice gauge theory and that the gKM model will have important implications for the knowledge of the zeta function at the same time.




In the present paper, we relax the condition imposed in \cite{matsuura2022kazakov} and consider the gKM model with two parameter degrees of freedom.
As mentioned above, while the Ihara zeta function counts only cycles without bumps (backtrackings and/or tails), general cycles on a graph typically have bumps. 
A graph zeta function that has been extended to count general cycles with bumps is known as the Bartholdi zeta function \cite{bartholdi2000counting} (see also \cite{mizuno2003bartholdi}), 
which has two parameters that count the length of the cycles and the number of bumps. 
The parameters of the gKM model considered in this paper can be related to these parameters. 
We show that the partition function of the gKM model in this general case is represented by the integral of the Bartholdi zeta function with unitary matrix weights on the edges.

The organization of this paper is as follows:
In Sect.~2, we introduce the terminology of the graph theory and review the basic properties of the Ihara and Bartholdi zeta functions. We also propose a matrix weighted Bartholdi zeta function. 
In Sect.~3, we introduce a generalized KM model on an arbitrary graph (gKM model). 
We show that the partition function of the gKM model is described by the integral of the weighted Bartholdi zeta function by unitary matrices on the edges. 
We explicitly write down the two representations of the partition function that arise by changing the order of integration over the two kinds of matrices; hermitian matrices on the vertices and the unitary matrices on the edges.
We also see the relation to the original KM model. 
In Sect.~4, we exactly evaluate the partition function of the gKM model on the cycle graph at finite $N$.
We see that the partition function is closely related to the generating function of the generalized Catalan numbers. 
In Sect.~5, we exactly evaluate the partition function of the model on an arbitrary graph at large $N$. 
The formulas developed in the previous paper \cite{matsuura2022kazakov} play essential roles. 
We show that the leading of the $1/N$ expansion of the free energy starts from the contribution of the zero area Wilson loops. 
We discuss the relation to the conventional large $N$ analysis of the original KM model. 
Sect.~6 is devoted to the conclusions and the discussions. 
In the appendix, we give proofs of Amitsur's theorem and an important identity used in Sect.~2.

\section{Graph zeta functions and their extensions}
\label{sec:Bartholdi zeta}


\subsection{Ihara and Bartholdi zeta function}
\label{sec:definition}

Let us begin with an explanation of graph theory terminology.
Suppose $G$ is a connected directed graph that has $n_V$ vertices
and $n_E$ edges.
Let us denote the set of vertices and edges as
$V$ and $E$, respectively.
A directed edge is written by a pair of vertices $e=\langle u,v\rangle$,
where $u=s(e)$ and $v=t(e)$ the ``source''  
and ``target'' of the edge arrow of $e$, respectively.
A reversed arrow edge for $e$ is called the inverse of $e$
and denoted by $e^{-1}=\langle v,u\rangle$.
We combine the set of the edges with their inverses as
\begin{equation}
E_D = \{\bse_a|a=1,\cdots,2n_E\} \equiv \{e_1,\cdots,e_{n_E},e^{-1}_1,\cdots,e^{-1}_{n_E}\}\,.
\end{equation}
In other words, $E_D$ is the direct sum of the set of edges $E$ and their inverses $E^{-1}$; $E_D=E\oplus E^{-1}$.

A path $P=(\bse_1,\cdots,\bse_k)$ $(\bse_a \in E_D)$ is a sequence of
the edges which satisfies $t(\bse_a)=s(\bse_{a+1})$ $(a=1,\cdots,k-1)$,
where $k$ is called the length of the path $P$ which is expressed as $|P|$.
If two paths $P=(\bse_1,\cdots,\bse_k)$ and $P'=(\bse'_1,\cdots,\bse'_l)$ satisfy
$t(\bse_k)=s(\bse'_1)$, we can construct a new path of length $k+l$
by connecting as
$PP'\equiv(\bse_1\cdots,\bse_k,\bse'_1,\cdots,\bse'_l)$.
A backtracking of $P$ is a part of $P$ which satisfies $\bse_{a+1}^{-1} = \bse_a$.

When a path $P=(\bse_1,\cdots,\bse_k)$ satisfies $s(\bse_1)=t(\bse_k)$, $P$ is called a cycle of length $k$.
A cycle $C$ is called primitive when $C$ satisfies $C \ne B^r$ $(r\ge 2)$ for any cycle $B$.
A cycle $C=(\bse_1,\cdots,\bse_k)$ is called tailless when $\bse_k^{-1}\ne \bse_1$, which is equivalent to that $C^2$ has no backtracking.
Backtracking or tail is also called bump,
and the number of bumps of a cycle $C$ is denoted by $b(C)$.

Two cycles $C=(\bse_1,\cdots,\bse_k)$ and $C'=(\bse'_1,\cdots,\bse'_k)$
are called equivalent when $\bse_a = \bse'_{a+r}$ for some integer $r$.
We denote the equivalence class including a cycle $C$ as $[C]$.
A cycle $C$ is called reduced when $C$ has neither backtracking nor tail.
We also denote the set of representatives of the equivalence classes of all kinds of cycles containing bumps by $[{\mathcal P}]$, 
and we denote the set of representatives of the equivalence classes of reduced cycles by $[{\mathcal P}_R] \subset [{\mathcal P}]$.

The Ihara zeta function \cite{Ihara:original,MR607504,sunada1986functions}
associated with a connected graph $G$ is defined as
\begin{equation}
  \zeta_G(q) = \prod_{C\in [{\cal P}_R]} \frac{1}{1-q^{|C|}}\,. 
  \label{eq:Ihara}
\end{equation}
Noting that, if $C$ is a primitive cycle, $C^{-1}$ is also a primitive cycle with the same length as $C$,
a set of the equivalence classes of the primitive cycles
of length $\ell$ can be decomposed
into
$\Pi_\ell^+ \sqcup \Pi_\ell^-$, where
$\Pi_\ell^-$ is the set of the inverse of the elements in $\Pi_\ell^+$ such that
$\Pi_\ell^-\equiv \left\{C^{-1}|C\in \Pi_\ell^+\right\}$.
Since these sets
$\Pi_\ell^+$ and $\Pi_\ell^-$ have the same number of elements by definition, namely $|\Pi_\ell^+|=|\Pi_\ell^-|$,
the Ihara zeta function \eqref{eq:Ihara} can be rewritten as
\begin{align}
  \zeta_G(q) = \prod_{\ell=1}^\infty \frac{1}{(1-q^\ell)^{2|\Pi^+_\ell|}}\,.
  \label{eq:Ihara2}
\end{align}
In the following, we call the element of $\Pi_\ell^+$ as
the equivalence class of {\em chiral primitive cycles} of length $\ell$
and write the set of the equivalence classes of all chiral primitive cycles as
\begin{equation}
  \Pi^+ \equiv \bigcup_{l=1}^\infty \Pi_l^+\,.
\end{equation}

We can further rewrite \eqref{eq:Ihara2} by using
the identity $(1-x)^{-1}=\exp(\sum_{m=1}^\infty \frac{x^m}{m})$.
Since any reduced cycle is represented by a positive power of a primitive cycle $C$ and the equivalence class $[C^n]$ $(n \in{\mathbb N})$ has $|C|$ different elements, the Ihara zeta function \eqref{eq:Ihara} can be regarded as a generating function of the number of reduced cycles;
\begin{align}
\zeta_G(q) = \exp\left( \sum_{k=1}^\infty \frac{N_k}{k} q^k \right)\,,
    \label{eq:Ihara zeta as generating function}
\end{align}
where $N_k$ is the number of the reduced cycles of length $k$.

Although we have considered only reduced cycles so far, 
we can define a function that counts up all cycles containing the bumps as
\begin{equation}
  \zeta_G(q,u) = \prod_{C\in [{\cal P}]} \frac{1}{1-q^{|C|} u^{b(C)}}\,,
  \label{eq:Bartholdi}
\end{equation}
where the product is taken over the equivalence class of all the cycles including bumps.
This function is called the Bartholdi zeta function associated with a connected graph $G$ \cite{bartholdi2000counting}%
\footnote{See also \cite{mizuno2003bartholdi} for a generalization.}.
Note that, if we take $u=0$, the Bartholdi zeta function reduces to
the Ihara zeta function;
\[
 \zeta_G(q,u=0)=\zeta_G(q)\,.
\]

It is remarkable that the Bartholdi zeta function (and thus also the Ihara zeta function) is represented as the inverse of a polynomial, even though the graph generally has infinitely many equivalence classes of cycles.
The key is the following  theorem \cite{amitsur1980characteristic,reutenauer1987formula} (see Appendix \ref{app:amitsur} for a proof):

\begin{description}
\item {\bf Theorem} (Amitsur).
{\em 
Let us consider ``letters'' $1,\cdots,k$ and call a sequence of the letters a word. 
In particular, we call such a word that cannot be written as a proper power of a shorter word ``primitive''. 
We call two words $w_1$ and $w_2$ equivalent when $w_2$ is obtained by a cyclic rotation of the letters of $w_1$.
We then denote the set of the representatives of the equivalent primitive words $L$. 
For square matrices $X_1,\cdots,X_k$, we define $X_w \equiv X_{i_1}\cdots X_{i_n}$ corresponding to a word $w$.  
Then the identity;
\begin{equation*}
\det \left(\bsone - X_1 - \cdots - X_k \right)
= \prod_{w \in L} \det\left(\bsone-X_w \right)
\end{equation*}
holds.}
\end{description}
In particular, the following corollary is useful:
\begin{description}
\item {\bf Corollary}.
{\em Let ${X}$ be a square matrix of the block form,
\[
{X}=\begin{pmatrix}
X_{11} & \cdots & X_{1k} \\
\vdots & \ddots & \vdots \\
X_{k1} & \cdots & X_{kk}
\end{pmatrix}\,,
\]
where $X_{ij}$'s are also square matrices of the same size $l$. 
For a word $w=i_1\cdots i_n \in L$, we define
\[
X_w = X_{i_1i_2}X_{i_2i_3}\cdots X_{i_{n-1}i_n} X_{i_n i_1}.
\]
Then the identity, 
\begin{equation*}
\det \left(\bsone_{kl} - X \right)
= \prod_{w\in L} \det\left(\bsone_l -X_w \right)\,,
\end{equation*}
holds.
}
\end{description}

To rewrite the Bartholdi zeta function \eqref{eq:Bartholdi} (and the Ihara zeta function \eqref{eq:Ihara}),
we define $2n_E\times 2n_E$ matrices $W$ and $J$ whose elements are defined by 
\begin{align}
    W_{\bse\bse'} = \begin{cases}
      1 & {\rm if}\ t(\bse) = s(\bse')\ {\rm and}\ \bse'^{-1}\ne \bse \\
      0 & {\rm others}
    \end{cases}\,,\quad
    J_{\bse\bse'} = \begin{cases}
      1 & {\rm if}\ \bse'^{-1}= \bse \\
      0 & {\rm others}
    \end{cases}\,,\quad
    \label{eq:WJ}
\end{align}
where $\bse, \bse' \in E_D$,
which are called the edge adjacency matrix and the bump matrix, respectively.
Using these matrices, we now consider a combination of the matrix $W$ and $J$
as $X\equiv q(W+uJ)$ whose elements are given by
\begin{equation*}
X_{\bse \bse'}=
\begin{cases}
  q W_{\bse \bse'} & \bse'\ne \bse^{-1} \\
  qu J_{\bse\bse'} & \bse'= \bse^{-1} \\
  0 & {\rm others}
\end{cases}\,.
\end{equation*}
For any cycle $C=(\bse_1,\cdots,\bse_k)$, which is not necessarily primitive, 
we can uniquely assign a sequence of the elements as 
$X_{C}\equiv X_{\bse_1\bse_2}\cdots  X_{\bse_{k-1}\bse_{k}}  X_{\bse_{k}\bse_{1}}=q^{|C|}u^{b(C)}$.
Since $X_{C}$ is common for the equivalence class of the cycle, 
by restricting the cycles to primitive cycles, we can write the Bartholdi zeta function as 
\begin{equation}
  \zeta_G(q,u)=\prod_{C\in[{\cal P}]} (1-X_{C})^{-1}\,. \nn
\end{equation}
Since we can identify the cycles with the words made of the edges, 
this is precisely the situation 
to
use (the corollary of) Amitsur's theorem.
We can therefore write the Bartholdi zeta function \eqref{eq:Bartholdi} as
the inverse of a polynomial of $q$ and $u$ as announced as 
\begin{equation}
\zeta_G(q,u) = \det \left(\bsone_{2n_E}- q(W+uJ)\right)^{-1}\,.
\label{eq:edge Bartholdi}
\end{equation}
Of course, by setting $u=0$, the same can be concluded for the Ihara zeta function \eqref{eq:Ihara}.

The expression \eqref{eq:edge Bartholdi} is sometimes called the edge Bartholdi zeta function because it is described through matrices $W$ and $J$ that characterize the relation among the edges of the graph.
Apart from this expression, there is another expression focusing on the relation among the vertexes; 
\begin{equation}
\zeta_G(q,u) = \bigl(\bsone_{n_V}-(1-u)^2q^2\bigr)^{-(n_E-n_V)}
\det\bigl(\bsone_{n_V}-q A + (1-u)q^2(D-(1-u) \bsone_{n_V})\bigr)^{-1}\,,
\label{eq:vertex Bartholdi}
\end{equation}
where $D$ is the diagonal matrix defined by
$D={\rm diag}_{v\in V}(\deg(v))$ and
the matrix $A$ is a square matrix of size $n_V$ called the vertex adjacency matrix
defined by
\begin{equation}
A_{vv'} = \sum_{\bse\in E_D}
\delta_{\langle{v},{v'}\rangle,\bse} \,.  \quad (v,v'\in V)
\label{eq:vertex A}
\end{equation}
In this paper, we will give a proof of the equivalence of \eqref{eq:edge Bartholdi} and \eqref{eq:vertex Bartholdi} as a corollary of the similar equivalence of the matrix weight Bartholdi zeta function introduced soon below.

\subsection{Matrix weighted Bartholdi zeta function}
\label{sec:matrix zeta}

We here consider to put an invertible $K\times K$ matrix $X_e$  to each edge $e$ of the graph.
We assume that the matrix on the inverse edge $e^{-1}$ is the inverse of the matrix on the edge $e$;
\begin{equation}
  X_{e^{-1}}=X_e^{-1}\,.
  \label{eq:matrix assumption}
\end{equation}
For a cycle $C=(\bse_{i_1}\cdots\bse_{i_n})$, 
we assign a matrix,
\[
X_C \equiv X_{\bse_{i_1}}\cdots X_{\bse_{i_n}}.
\]
We then propose an extension of the Bartholdi zeta function, 
\begin{align}
  \zeta_G(q,u;X) &\equiv \prod_{C\in [{\cal P}]} \det\bigl( \bsone_{K} - q^{|C|} u^{b(C)} X_C \bigr)^{-1}\,, 
  \label{eq:W-Bartholdi}
\end{align}
which we call the matrix weighted Bartholdi zeta function
in the following. 
The matrix weighted Bartholdi zeta function \eqref{eq:W-Bartholdi} is an extension of the weighted Bartholdi zeta function defined in \cite{sato2006weighted,choe2007bartholdi} where the weights on the edges are supposed to be c-numbers.
Note that it becomes the matrix weighted Ihara zeta function introduced in \cite{matsuura2022kazakov} by taking $u=0$.

Suppose a cycle with backtracking as $\tC=P_1 \bse \bse^{-1} P_2$.
Then the matrix $X_\tC$ reduces as
$X_\tC=X_{P_1} X_{\bse} X_{\bse^{-1}} X_{P_2} = X_{P_1}X_{P_2}$ because we have assumed $X_{\bse^{-1}}=X_\bse^{-1}$. 
The same reduction occurs also for tails when $X_\tC$ is included in the determinant as \eqref{eq:W-Bartholdi}. Repeating this reduction, the matrix $X_{\tC}$ finally reduces to the matrix associated with a reduced cycle. 
In general, a reduced cycle is a positive power of a primitive reduced cycle $C$.
We then denote the set of representatives of primitive cycles that are equivalent to $C^k$ $(k\in\N)$ after eliminating the bumps by $[{\cal B}(C^k)] \subset [{\cal P}]$. 
We also denote the set of the representatives of primitive cycles that reduce to a point (vertex) by eliminating the bumps by $[{\cal B}_0] \subset [{\cal P}]$.
From the consideration above,
$X_\tC$ for $\tC\in [{\cal B}(C^k)]$ reduces to 
$X_C^k$ and thus we can rewrite \eqref{eq:W-Bartholdi} as
\begin{align}
  \zeta_G(q,u;X) &= {\cal V}_G(q,u)^K
  \prod_{C\in [{\cal P}_R]}
  \prod_{k=1}^\infty
  \prod_{\tC\in [{\cal B}(C^k)]}
  \det\bigl(\bsone_K- q^{|\tC|}u^{b(\tC)}X_{C}^k \bigr)^{-1}\,,
  \label{eq:med}
\end{align}
where 
\begin{equation}
{\cal V}_G(q,u) \equiv \prod_{\tC\in [{\cal B}_0]} \frac{1}{1-q^{|\tC|}u^{b(\tC)}}\,.
\label{eq:VG}
\end{equation}
We can further evaluate it as
\begin{align}
   \prod_{k=1}^\infty
  \prod_{\tC\in [\cB(C^k)]}
  \det\left(\bsone_K- q^{|\tC|}u^{b(\tC)}X_{C}^k \right)^{-1}
  &= \exp\left\{
  \sum_{k=1}^\infty \sum_{m=1}^\infty \frac{1}{m}
  \left(\sum_{\tC\in [\cB(C^k)]} (q^{|\tC|}u^{b(\tC)})^m \right)
  \Tr(X_C^{km})
  \right\} \nn \\
  &= \exp\left\{
  \sum_{n=1}^\infty \frac{1}{n}
    \sum_{k|n} \left( k \sum_{\tC\in [\cB(C^k)]} (q^{|\tC|}u^{b(\tC)})^{n/k} \right)
  \Tr(X_C^{n}) \right\} \nn \\
  &= \exp\left\{
  \sum_{n=1}^\infty \frac{1}{n}
   \left(\sum_{\tC\in [\cB(C)]} q^{|\tC|}u^{b(\tC)}\right)^n
    \Tr(X_C^{n}) \right\}\,, \nn
\end{align}
where we have used the identity,
\begin{align}
  \sum_{k|n} \left( k \sum_{\tC\in [\cB(C^k)]} (q^{|\tC|}u^{b(\tC)})^{n/k} \right)
  = \left(\sum_{\tC\in [\cB(C)]} q^{|\tC|}u^{b(\tC)}\right)^n\,,
  \label{eq:id fC}
\end{align}
which is proven in Appendix \ref{app:fC}.
Therefore the matrix weighted Bartholdi zeta function is finally written as
\begin{align}
  \zeta_G(q,u;X) &= {\cal V}_G(q,u)^K
  \prod_{C\in [{\cal P}_R]}
  \exp\left(
    \sum_{n=1}^\infty \frac{f_{C}(q,u)^n}{n}
     \Tr(X_C^{n}) \right)\,,
     \label{eq:path W-Bartholdi}
\end{align}
where
\begin{equation}
  f_{C}(q,u) \equiv \sum_{\tC\in [\cB(C)]} q^{|\tC|}u^{b(\tC)}\,.
  \label{eq:FC}
\end{equation}
This expression plays an essential role in evaluating the partition function
of the KM model in the following section.

We note that the matrix weighted Bartholdi zeta function can be further rewritten as
\begin{equation}
  \zeta_G(q,u;X) = {\cal V}_G(q,u)^K
  \prod_{C\in [{\cal P}_R]}
  \det(\bsone_K-f_C(q,u)X_C)^{-1}\,,\nn
\end{equation}
which can be regarded as an extension of the Ihara zeta function in the sense that we count rather $f_C(q,u)X_C$ not $q^{|C|}$ for a primitive reduced cycle $C$.

\subsection{Determinant expressions of the matrix weighted Bartholdi zeta function}
\label{sec:edge zeta}

As same as the original Bartholdi zeta function,
the matrix weighted Bartholdi zeta function \eqref{eq:W-Bartholdi} can be expressed as the inverse of the determinant.
This is achieved by extending the edge adjacency matrix
and the bump matrix \eqref{eq:WJ} as
\begin{align}
    (W_X)_{\bse\bse'} = \begin{cases}
    X_{\bse} & {\rm if}\ t(\bse) = s(\bse')\ {\rm and}\ \bse'^{-1}\ne \bse \\
      0 & {\rm others}
    \end{cases}\,,\quad
    (J_X)_{\bse\bse'} = \begin{cases}
      X_{\bse} & {\rm if}\ \bse'^{-1}= \bse \\
      0 & {\rm others}
    \end{cases}\,.
    \label{eq:WJw}
\end{align}
Repeating the discussion above the equation \eqref{eq:edge Bartholdi}, we can show that the matrix weighted Bartholdi zeta function can be written as
\begin{align}
  \zeta_G(q,u;X)
  &= \det\bigl( \bsone_{2Kn_E} - q(W_X + u J_X)\bigr)^{-1}\,,
  \label{eq:edge W-Bartholdi}
\end{align}
as a direct result of Amitsur's theorem.

We can further show that, as same as \eqref{eq:vertex Bartholdi}, the matrix weighted Bartholdi zeta function can be expressed through the matrix weighted vertex adjacency matrix of the size $K n_V$,
\begin{equation}
(A_X)_{vv'} = \sum_{\bse\in E_D} X_{\bse}\,
 \delta_{\langle{v},{v'}\rangle,\bse} \,,
 \label{eq:vertex Amat}
\end{equation}
as
\begin{equation}
\zeta_G(q,u;X) = \bigl(1-(1-u)^2q^2\bigr)^{-K(n_E-n_V)}
\det\bigl(\bsone_{Kn_V}-q A_X + (1-u)q^2(D-(1-u) \bsone_{Kn_V})\bigr)^{-1}\,,
\label{eq:vertex W-Bartholdi}
\end{equation}
where $D$ has been redefined as
\begin{equation}
  D_{vv'} = \deg(v) \delta_{vv'} \bsone_K\,.
  \label{eq:matD}
\end{equation}

Let us prove it 
by using the strategy described in \cite{bass1992ihara}
(see also \cite{matsuura2022kazakov}):
Firstly, we define matrices $S_X$ and $T_X$ of size $n_E\times n_V$
whose elements are square matrices of size $K$;
\begin{align}
    (S_X)_{ev} \equiv \begin{cases}
      X_e^{-1} & {\rm if}\ v=s(e) \\
      0 & {\rm others}
    \end{cases}\,,\quad
    (T_X)_{ev} \equiv \begin{cases}
      X_e & {\rm if}\ v=t(e) \\
      0 & {\rm others}
    \end{cases}\,.
    \nn
\end{align}
We also define $S\equiv (S_X)|_{X=1}$ and $T\equiv (T_X)|_{X=1}$.
Then we can easily show
\begin{align}
    S^T T_X  + T^T S_X = A_X, \quad
    S^T S + T^T T = D\,.
    \nn
\end{align}
Using these matrices, we define
\begin{align*}
    L&\equiv \begin{pmatrix}
        \bs{1}_{Kn_V} & qS^T & q T^T \\
        -tq S+T_X & (1-t^2q^2) \bs{1}_{Kn_E}  & 0 \\
        -tq T+S_X & 0 & (1-t^2q^2) \bs{1}_{Kn_E}
    \end{pmatrix}\,, \quad
    M\equiv \begin{pmatrix}
        (1-t^2q^2) \bs{1}_{Kn_V} & 0 & 0 \\
        tqS-T_X & \bs{1}_{Kn_E} & 0 \\
        tqT-S_X & 0 & \bs{1}_{Kn_E}
    \end{pmatrix}\,.
\end{align*}
After a straightforward calculation, we obtain
\begin{align*}
    LM &= \begin{pmatrix}
    \bsone_{Kn_V}-q A_X + tq^2(D-t \bsone_{Kn_V}) & qS^T & qT^T \\
        0 & (1-t^2q^2)\bs{1}_{Kn_E} & 0 \\
        0 & 0 & (1-t^2 q^2) \bs{1}_{Kn_E}
    \end{pmatrix}\,,  \nn\\
    ML&= \begin{pmatrix}
        (1-t^2q^2)\bs{1}_{Kn_V} &  q(1-t^2q^2)S^T \hspace{2cm} q(1-t^2q^2)T^T \\
        \begin{array}{c} 0 \\ 0 \end{array} &
        \left(\bs{1}_{2Kn_E}-q(W_X+(1-t)J_X)\right)
        \left(\bs{1}_{2Kn_E}-tq J_X\right)
    \end{pmatrix}\,.
    \nn
\end{align*}
Using $\det(LM)=\det(ML)$ and
${\rm det}\left( \bs{1}_{2Kn_E}-tq J_X \right) = (1-t^2 q^2)^{Kn_E}$,
we find
\begin{multline}
\qquad
(1-t^2 q^2)^{2 K n_E}\det\left(
\bsone_{Kn_V}-q A_X + tq^2 (D-t \bsone_{Kn_V})
\right)\\
=(1-t^2q^2)^{K (n_V+n_E)}\det\left(\bs{1}_{2Kn_E}-q(W_X+(1-t)J_X)\right)\, ,
\qquad
\nn
\end{multline}
and can conclude (\ref{eq:vertex Bartholdi}) by setting $t=1-u$. 
Note that the assumption \eqref{eq:matrix assumption} is required in this computation. 

\section{Graph Kazakov-Migdal model and Bartholdi zeta function}
\label{sec:KM model}

\subsection{The partition function of the graph Kazakov-Migdal model as Bartholdi zeta function}
\label{sec:gKM}

The KM model \cite{kazakov1993induced} is a lattice gauge theory
with the gauge group $SU(N)$ which is
originally defined on the $D$-dimensional square lattice with the action \eqref{eq:SKM}. 
We can naturally extend it to a model on the graph.
Suppose that a Hermitian matrix $\Phi_v$ and a unitary matrix $U_e \in U(N)$ live
on each vertex $v\in V$ and each edge $e\in E$ of a given graph $G$, respectively.
If we consider gauge invariant operators of quadratic in $\Phi$ as in the original KM model, we can consider $\Tr\Phi_v^2$ for each vertex $v$ and $\Tr \Phi_{s(e)}U_e \Phi_{t(e)} U_e ^\dagger$ 
and $\Tr\left(\Phi_{s(e)}^2+\Phi_{t(e)}^2\right)$ for each edge $e$. Therefore, if we assign global coupling constants to these operators, the most general action is given by 
\begin{align}
S
&=\Tr \left\{
\frac{m_0^2}{2} \sum_{v\in V}\Phi_v^2
+ q\sum_{e \in E}
\left(
\frac{r}{2}\left(\Phi_{s(e)}^2+\Phi_{t(e)}^2\right)
-\Phi_{s(e)}U_e \Phi_{t(e)}U_e^\dag
\right)
\right\} \nn \\
&=\Tr \left\{
\frac{1}{2}\sum_{v\in V}
\left(m_0^2 + qr \deg v \right)
\Phi_v^2
- q\sum_{e\in E} \Tr \Phi_{s(e)}U_e \Phi_{t(e)}U_e^\dag
\right\}\,. 
\label{eq:Sgeneral}
\end{align}
In the following, we fix the parameters $r$ and $m_0^2$ by using an additional constant $u$ as $r=q(1-u)$ and $m_0^2=1-q^2(1-u)^2$, 
\begin{equation}
S_{\rm gKM} =
\Tr \left\{
\frac{1}{2}\sum_{v\in V}
(1-q^2(1-u)^2+q^2(1-u) \deg v)
\Phi_v^2
- q\sum_{e \in E}
\Phi_{s(e)}U_e \Phi_{t(e)}U_e^\dag
\right\}\,,
\label{eq:SgKM}
\end{equation}
in order to see a connection to the Bartholdi zeta function soon later. 
Although this parametrization is not the most general one, the action \eqref{eq:SgKM} reproduces the action of the original KM model \eqref{eq:SKM} by setting the graph $G$ to the $D$-dimensional square lattice and tuning $q$ and $u$ appropriately.  
We call the model with the action \eqref{eq:SgKM} the graph Kazakov-Migdal (gKM) model.
Note that the model considered in \cite{matsuura2022kazakov} is obtained by setting $u=0$.

We next consider the partition function of the gKM model,
\begin{equation}
  Z_{\rm gKM} = \int \prod_{v\in V}d\Phi_v \prod_{e\in E} dU_e
  \, e^{-\beta S_{\rm gKM}}\,.
  \label{eq:ZgKM}
\end{equation}
As discussed in \cite{kazakov1993induced} (see also \cite{matsuura2022kazakov}), 
there are two different representations of the partition function $Z_{\rm gKM}$
depending on whether the scalar fields $\Phi_v$ is integrated first or the gauge fields $U_e$ is integrated first.
Let us first consider the case of integrating the scalar fields $\Phi_v$ first.
The action \eqref{eq:SgKM} is bilinear in the scalar fields $\Phi_v$ and
can be written as
\begin{equation}
  S_{\rm gKM} =
  \frac{1}{2} \Phi_{v,a}
  \left(\bsone_{n_V N^2} -q A_U
  + q^2(1-u)(D-(1-u)\bsone_{n_V N^2}) \right)_{va,v'b}
  \Phi_{v'b}\,,
\end{equation}
where $a,b=1,\cdots,N^2$ and $A_U$ is the adjacency matrix given by \eqref{eq:vertex Amat} with $X_e=U_e\otimes U_e^\dagger$
and $D$ is the degree matrix defined in \eqref{eq:matD}.
The matrix that appeared in this expression is the same as the one that appeared in the vertex representation of the matrix weighted Bartholdi zeta function \eqref{eq:vertex W-Bartholdi}.
Therefore the integration over the scalar fields $\Phi_v$ in the partition function yields
\begin{align}
  Z_{\rm gKM} &= \left(\frac{2\pi}{\beta}\right)^{\frac{1}{2}n_V N^2}\int \prod_{e\in E} dU_e\,
{\det}\lrpar{\bsone_{n_V N^2} -q A_U
+ q^2(1-u)(D-(1-u)\bsone_{n_V})}^{-\frac{1}{2}} \nn \\
  &= \left(\frac{2\pi}{\beta}\right)^{\frac{1}{2}n_V N^2}
  (1-(1-u)^2q^2)^{\frac{1}{2}{(n_E-n_V)N^2}}\int \prod_{e\in E} dU_e\, \zeta_G(q,u;U)^\frac{1}{2} \nn \\
  &= \left(\frac{2\pi}{\beta}\right)^{\frac{1}{2}n_V N^2}
  (1-(1-u)^2q^2)^{\frac{1}{2}{(n_E-n_V)N^2}}
  {\cal V}_G(q,u)^{\frac{N^2}{2}} \nn \\
  &\hspace{4cm}\times
  \int\prod_{e\in E} dU_e\,
      \prod_{C\in\Pi^+} e^{\sum_{m=1}^\infty \frac{1}{m}f_C(q,u)^m |\Tr P_C(U)^m|^2}
  \label{eq:Z_zeta}
\end{align}
where $\zeta_G(q,u;U)$ is the matrix weighted Bartholdi zeta function \eqref{eq:W-Bartholdi} with $X_e=U_e\otimes U_e^\dagger$ 
and we have used \eqref{eq:path W-Bartholdi} 
in the last line.
We note that, using the left-right invariance of the Haar measure $\prod_{e\in E}dU_e$ in \eqref{eq:ZgKM} and the gauge invariance of the action \eqref{eq:SgKM}, 
we can fix $U_e=1$ on the edges of a spanning tree of the graph $G$, that is, a subgraph of $G$ that contains all the vertices of $G$ and has no cycles.


\subsection{The dual description of the partition function and a relation to covering graphs}
\label{sec:IZrep}

We next consider integrating over the gauge fields $U_e$ first.
To this end, we need the so-called Harish-Chandra-Itzykson-Zuber (HCIZ) integral formula \cite{itzykson1980planar};
\begin{equation}
  \int dU \, e^{t \Tr AUBU^\dagger}
  = \frac{G(N+1)}{t^\frac{N^2-N}{2}}
  \frac{{\det}_{i,j} \lrpar{e^{t a_i b_j}}} {\Delta(a)\Delta(b)}\,,
  \label{eq:IZ}
\end{equation}
where
$U$ is a unitary matrix of size $N$,
$dU$ is the Haar measure of $U(N)$ normalized as $\int dU = 1$,
$G(N+1)=\prod_{i=1}^{N-1} i!$ is the Barnes double gamma function,
$A$ and $B$ are Hermitian matrices whose eigenvalues are $(a_1,\cdots,a_N)$
and $(b_1,\cdots,b_N)$, respectively,
and $\Delta(a)$ and $\Delta(b)$ are the Vandermonde determinant with respect to $A$ and $B$,
respectively;
\[
  \Delta(a) = \prod_{1\le i<j\le N} (a_j-a_i)\,, \quad
  \Delta(b) = \prod_{1\le i<j\le N} (b_j-b_i)\,.
\]

We apply the formula \eqref{eq:IZ} to the partition function of the gKM
model \eqref{eq:ZgKM}.
Since the integrand includes only the eigenvalues of
the Hermitian matrices $\Phi_v$, we also use a mapping
from the matrix integral to the
integral over the eigenvalues
(see e.g.\cite{marino2004houches}),
\[
  \int d\Phi f(\Phi) =
  \frac{(2\pi)^{\frac{N(N-1)}{2}}}{G(N+2)} \int \prod_{i=1}^N d\phi_i\,
  \Delta(\phi)^2 f(\phi)\,,
\]
for a Hermitian matrix $\Phi$ with eigenvalues $\phi_i$ ($i=1,\cdots,N$).
We then obtain
\begin{align}
  Z_{\rm gKM}
  &={\cal N}  \int \lrpar{ \prod_{v\in V}\prod_{i=1}^N d\phi_{v,i} }
  \lrpar{\prod_{v'\in V}\Delta(\phi_{v'})^{2-\deg v'}}
  \sum_{\sigma_1,\cdots,\sigma_{n_E}\in S_N} \sgn(\sigma_1\cdots \sigma_{n_E})
  e^{-\frac{1}{2} \phi_{v,i}(D_\sigma)_{vi;v'j}\phi_{v',j}}\,, 
  \label{eq:IZrep}
\end{align}
where 
\begin{equation}
  {\cal N} \equiv \frac{{G(N+1)}^{n_E-n_V}} {(N!)^{n_V}}
  \frac{(2\pi)^{\frac{N(N-1)}{2}n_V}}
       {\beta^{\frac{N^2}{2}n_V} q^{\frac{N(N-1)}{2}n_E}}\,, \nn
\end{equation}
and 
\begin{equation}
  (D_\sigma)_{vi;v'j} \equiv \Bigl( 1+(1-u)q^2(\deg v - (1-u) ) \Bigr) \delta_{vv'}\delta_{ij}
  -q A_{vv'}\Bigl( \delta_{i,\sigma_{\langle vv'\rangle}^{-1}(j)} + \delta_{j,\sigma_{\langle vv'\rangle}(i)} \Bigr)\,,
  \label{eq:Dsigma}
\end{equation}
with the vertex adjacency matrix $A$ given in \eqref{eq:vertex A}.

It is remarkable that the matrix \eqref{eq:Dsigma} is the matrix appeared in the vertex representation of the Bartholdi zeta function \eqref{eq:vertex Bartholdi}
corresponding to a covering graph of $G$: 
Imagine that $N$ copies of the graphs $G$ are stacked with their vertices aligned. 
We label the vertex $v$ of the $i$'s layer as $(v,i)$. 
Let $\langle v,v'\rangle$ be an edge of the original graph $G$, 
and we consider to reconnect the $N$ vertices above $v$ and $v'$ by edges in a one-to-one relationship.
If we connect the vertices in the same layer $(v,i)$ and $(v',i)$ ($i=1,\cdots,N$) for all $\langle v,v'\rangle\in E$, 
it becomes simply $N$ copies of the original graph $G$. 
Instead, we can connect vertices in different layers, $(v,i)$ and $(v',j)$ $(i\ne j)$.
In this case, by making such recombination on every edge, 
we can create a graph that is different from $G$, which is called a covering graph of $G$.
Note that the obtained covering graph is disconnected in general. 
Since which layers are connected at edge $e$ is specified by the permutation of $N$, 
the vertex adjacency matrix of the covering graph is expressed as a direct product of the vertex adjacency matrix $A$ of the original graph and elements of $S_N$. 
It is exactly the one that appeared in the second term of \eqref{eq:Dsigma}. 
From the construction, the degree of the vertex $(v,i)$ of the covering graph is the same with the degree $v$ of the original graph $G$. 
Therefore we see that \eqref{eq:Dsigma} is exactly the matrix that appeared in the vertex representation of the Bartholdi zeta function \eqref{eq:vertex Bartholdi} for the covering graph.

We note that there is a residual gauge symmetry in the expression \eqref{eq:IZrep} to permute the eigenvalues of $\Phi_v$ as $\phi_{v,i}\to\phi_{v,\tau_v(i)}$ by ${}^\forall \tau_v\in S_N$, 
which transforms the permutation $\sigma_e$ on the edge $e$ as $\sigma_e\to\tau_{s(e)}\sigma_e\tau_{t(e)}$.
Therefore, we can fix this gauge by setting the permutations on the edges of a spanning tree of the graph as $\sigma_e=1$.
This symmetry corresponds to the freedom to swap permutations that create the same covering graph.
We will see it in a concrete example in the next section. 

\subsection{Graph Kazakov-Migdal model on a $d$-regular graph}
\label{sec:regular}

We next restrict the graph to a $d$-regular graph, that is, a graph where each vertex has the same degree $d$. 
In this case, the action of the gKM model \eqref{eq:Sgeneral} can be written in the same manner with the original KM model \eqref{eq:SKM} as 
\begin{equation}
  S_{\rm gKM}=q\Tr\left\{
  \frac{m^2}{2}\sum_{v\in V}\Phi_v^2
  - \sum_{e\in E} \Phi_{s(e)}U_e\Phi_{t(e)}U_e^\dagger
  \right\}\,,
  \label{eq:SgKM reguar}
\end{equation}
where the mass parameter $m^2$ is related to the parameters $q$ and $u$ as 
\begin{equation}
  m^2 = q^{-1} + q\bigl( (1-u)d - (1-u)^2 \bigr)\,.
  \label{eq:mass}
\end{equation}
Note that, although $q$ can be absorbed into $\beta$ in the definition of the partition function \eqref{eq:ZgKM}, 
we have kept it to make the relation to the original parametrization clearer. 

\begin{figure}[H]
\begin{center}
\subcaptionbox{The zero area Wilson loop on a cycle in $\cB_0$.}[.45\textwidth]{
\includegraphics[scale=0.5]{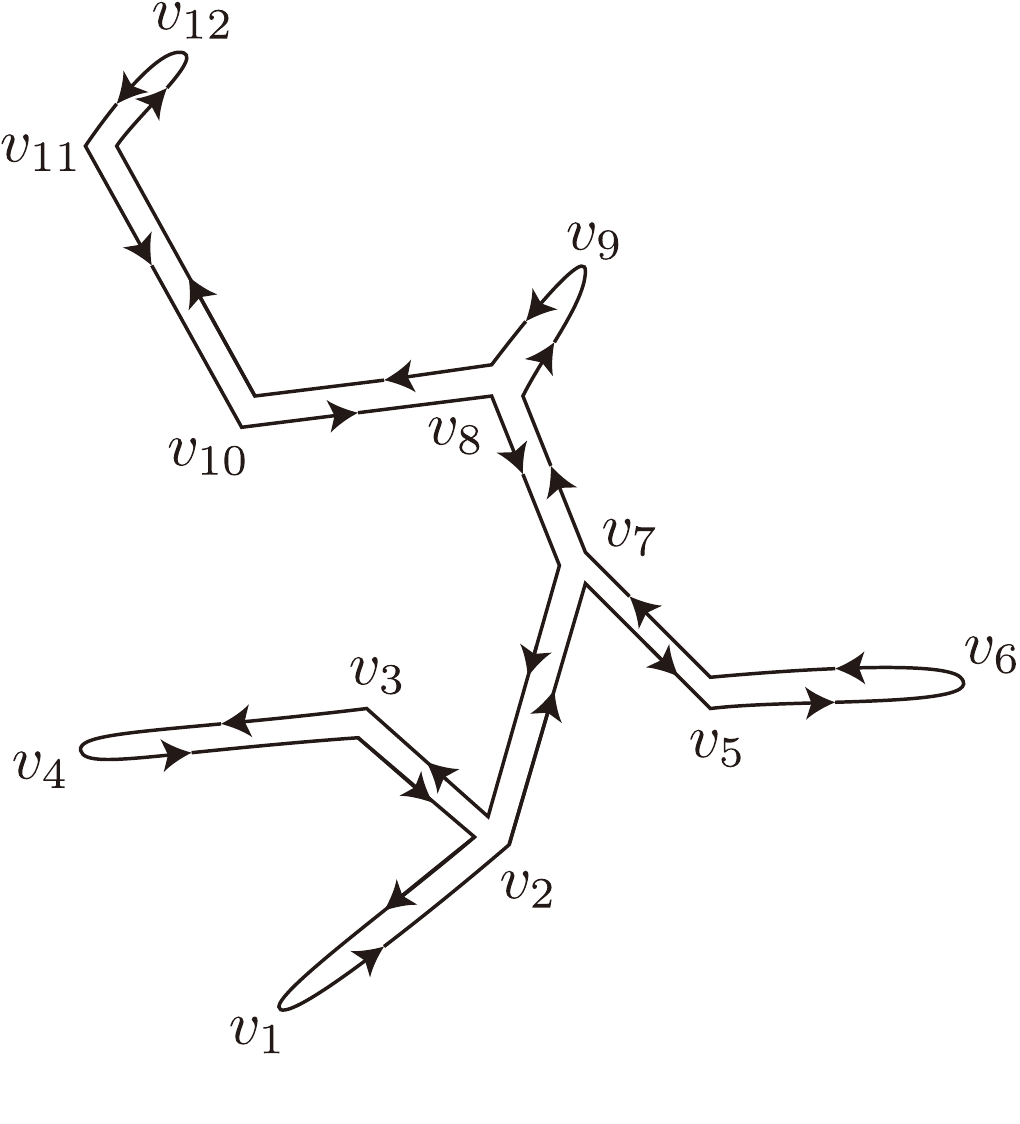}
}\qquad  
\subcaptionbox{The non-zero area Wilson loop on a cycle in $\cB(C)$.}[.45\textwidth]{
\includegraphics[scale=0.5]{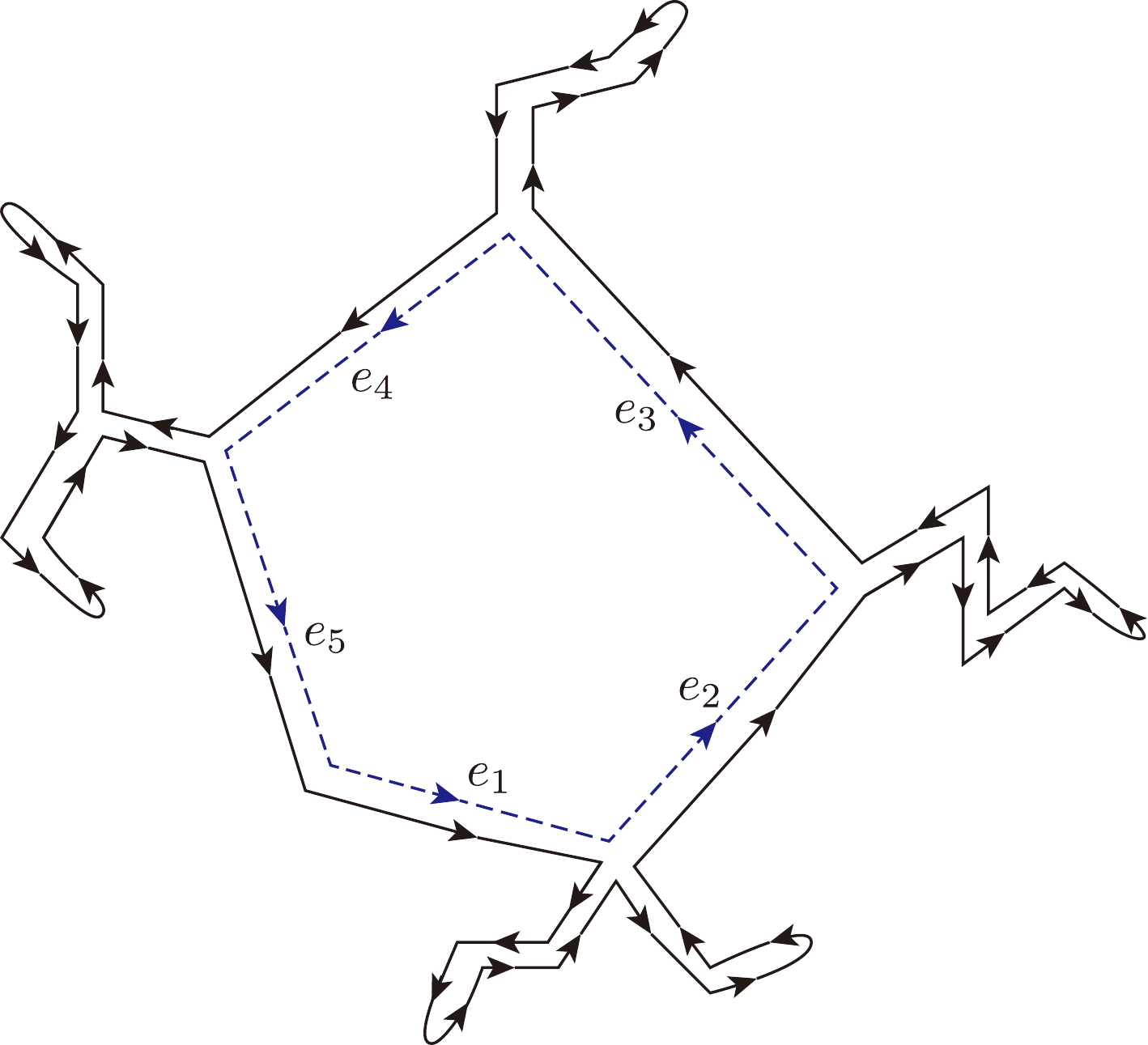}
}
\end{center}
\caption{
Examples of the Wilson loops on the cycles with branches and
bumps.
The Wilson loop on the cycle (a)
reduces to a point (vertex)
and $P_C(U)=1$
since $UU^\dag=1$ on round trips over the edges.
For the same reason,
the Wilson loop on the cycle (b)
gives the same contribution to
the Wilson loop on
the cycle $C=(e_1,e_2,e_3,e_4,e_5)$
without branches and bumps
(dashed cycle).
}
\label{Branched Wilson loops}
\end{figure}

In this parametrization, we can write the partition function as 
\begin{align}
  Z_{\rm gKM} &= 
  \left(\frac{2\pi}{\beta q\, m^2}\right)^{\frac{1}{2}n_V N^2}\int \prod_{e\in E} dU_e\,
{\det}\lrpar{\bsone_{n_V N^2} - m^{-2} A_U}^{-\frac{1}{2}} \nn \\
&= \left(\frac{2\pi}{\beta q\, m^2}\right)^{\frac{1}{2}n_V N^2}\int \prod_{e\in E} dU_e\,
\exp \left( \frac{1}{2}
\sum_{k=1}^{\infty} \frac{m^{-2k}}{k} \Tr A_U^k
\right) \nn \\
&= \left(\frac{2\pi}{\beta q\, m^2}\right)^{\frac{1}{2}n_V N^2}\int \prod_{e\in E} dU_e\,
\exp \left( \frac{1}{2}
\sum_{k=1}^{\infty} \frac{m^{-2k}}{k} \sum_{|C|=k}
\bigl| \Tr P_C(U) \bigr|^2
\right)\,,
\label{eq:Ztochu}
\end{align}
where we have used the fact that $\Tr A_U^k$ counts all Wilson loops with length $k$ and the summation of the cycles in the parenthesis of the last line runs over all cycles with length $k$. 
Since any cycle can be written as a positive power of a primitive cycle 
and 
the set of the representatives of primitive cycles $[{\cal P}]$ can be decomposed into
$[\cB_0]$ and $[\cB({C}^j)]$ (${C}\in [{\cal P}_R]$, $j\in\N$),
we can estimate the expression in the parenthesis as
(see also Fig.~\ref{Branched Wilson loops}.)
\begin{align}
  \frac{1}{2}
&\sum_{k=1}^{\infty} \frac{m^{-2k}}{k}  \sum_{|C|=k}
\bigl| \Tr P_C(U) \bigr|^2 \nn \\
&= \frac{1}{2}\sum_{C\in [{\cal P}]}
|C| \sum_{l=1}^\infty \frac{m^{-2l|C|}}{l|C|} 
\bigl| \Tr P_C(U)^l \bigr|^2 \nn \\
&= \frac{N^2}{2}\sum_{\tC\in [\cB_0]} \sum_{l=1}^\infty \frac{m^{-2l|\tC|}}{l}
+ \sum_{{C}\in [{\cal P}_R]} 
\sum_{j=1}^\infty \sum_{\tC\in[\cB({C}^j)]}
\sum_{l=1}^\infty \frac{m^{-2l|\tC|}}{l} 
\bigl| \Tr P_{{C}}(U)^{jl} \bigr|^2 \nn \\
&= -\frac{N^2}{2} \sum_{\tC\in [\cB_0]} \log\left(1-m^{-2|\tC|}\right)
+\sum_{{C}\in[{\cal P}_R]} 
\sum_{n=1}^\infty \frac{1}{n} \left(
\sum_{\tC\in[\cB({C})]} m^{-2|\tC|}
\right)^n
\bigl| \Tr P_{{C}}(U)^{n} \bigr|^2\,, \nn
\end{align}
where we have used the fact that the equivalence class of the cycle $C^l$ $(C\in[{\cal P}])$ has $|C|$ different elements in the first line 
and 
we have repeated the same calculation above \eqref{eq:id fC} to obtain the last line. 
We can then rewrite \eqref{eq:Ztochu} as 
\begin{align}
  Z_{\rm gKM}&= 
  \left(\frac{2\pi}{\beta q\, m^2}\right)^{\frac{1}{2}n_V N^2}
  {\cal V}_G(m^{-2},1)^{\frac{N^2}{2}}
  \int \prod_{e\in E} dU_e\,
  \exp\left(
  \sum_{{C}\in[\Pi^+]} 
  \sum_{n=1}^\infty \frac{1}{n} f_{{C}}(m^{-2},1)^n
  \bigl| \Tr P_{{C}}(U)^{n} \bigr|^2
  \right)\,,
  \label{eq:ZgKM_reular}
\end{align}
where ${\cal V}_G(q,u)$ and $f_C(q,u)$ are defined in \eqref{eq:VG} and \eqref{eq:FC}, respectively.

Comparing this expression to \eqref{eq:Z_zeta}, we obtain non-trivial identities satisfied by $d$-regular graphs:
\begin{align}
\begin{split}
    (qm^2)^{-n_V} {\cal V}_G(m^{-2},1) &=  (1-(1-u)^2q^2)^{n_E-n_V} {\cal V}_G(q,u)\,, \\
    f_C( m^{-2}, 1) &= f_C(q,u) \,.
\end{split}
\label{eq:formulae1}
\end{align}
In particular the relation ${\cal V}_G(q,0)=1$ and $f_C(q,0)=q^{|C|}$  yields the identities, 
\begin{align}
\begin{split}
  \prod_{\tC\in[\cB_0]} \left(1-(q^{-1}+(d-1)q)^{|\tC|}\right)^{-1} 
  &=\bigl( (1+(d-1)q^2) (1-q^2)^{d-2} \bigr)^{n_V}
  \,, \\
  \sum_{\tC\in[\cB(C)]}(q^{-1}+(d-1)q)^{-|\tC|} &=  q^{|C|}  \,.
\end{split}
\label{eq:formulae2}
\end{align}

\section{Exact partition function of the graph Kazakov-Migdal model on cycle graphs at finite $N$}
\label{sec:Cn}

In \cite{matsuura2022kazakov}, the integral \eqref{eq:IZrep} with $u=0$ for the cycle graph $C_n$,
that is, a connected graph that has $n$ vertices and $n$ edges and $\deg v =2$ for ${}^\forall v\in V$, 
was evaluated exactly at finite $N$. 
In this section, we show that the same can be performed for the gKM model in general parameters.

\begin{figure}[H]
\begin{center}
\includegraphics[scale=0.4]{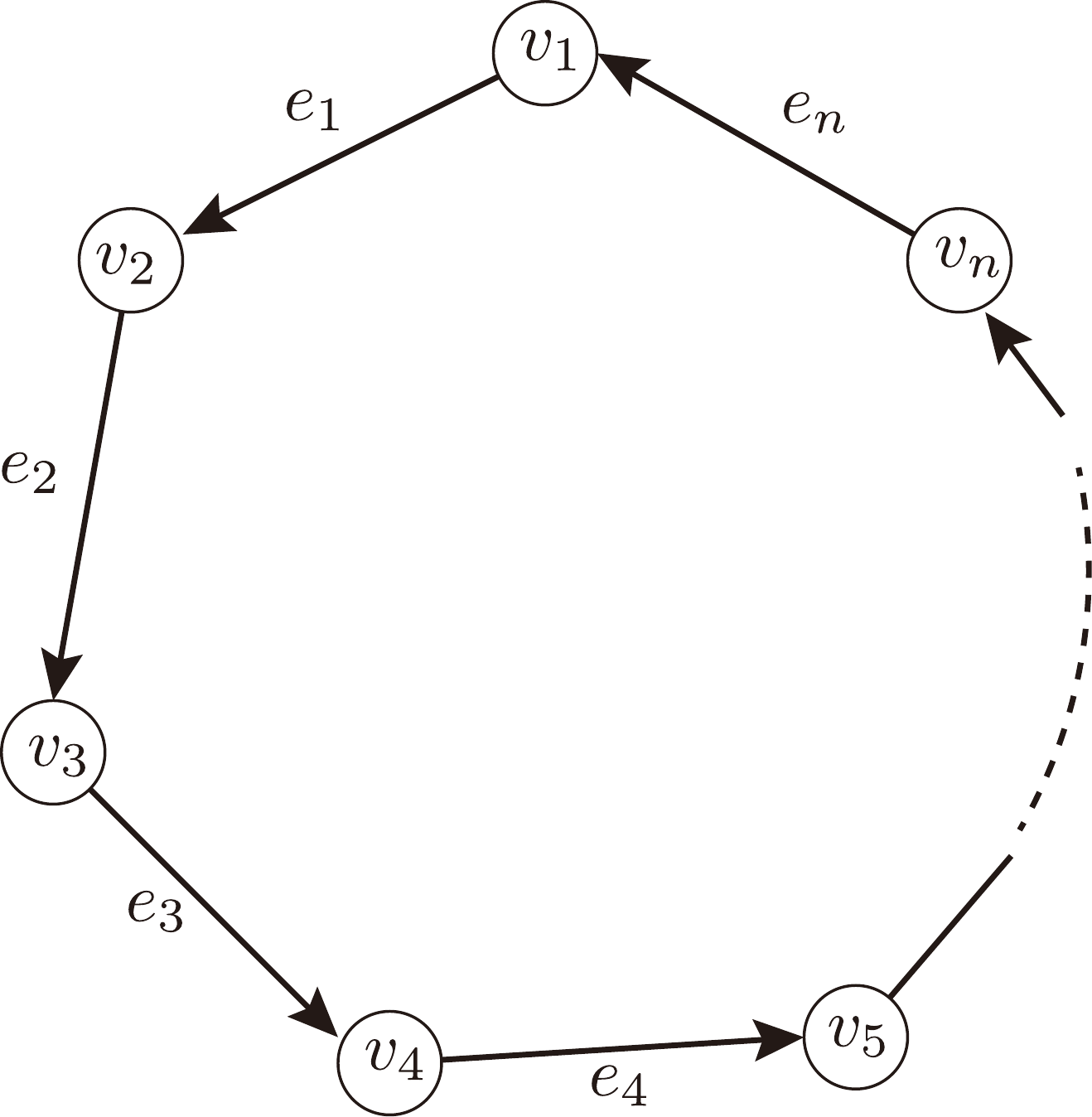}
\end{center}
\caption{A cycle graph $C_n$ with $n$ vertices and edges}
\label{fig:Cn}
\end{figure}

\subsection{Evaluation by unitary matrix integral}

Let us first evaluate the Bartholdi zeta function for $C_n$ using the expression \eqref{eq:vertex Bartholdi}; 
\begin{align}
  \zeta_{C_n}(q,u)=(\det M_n)^{-1}\,, 
\end{align}
where the matrix $M_n$ is defined by 
\begin{align}
  M_1 &= 1-2q+(1-u^2)q^2, \quad 
  M_2 = \begin{pmatrix}
  1+(1-u^2)q^2 & -2q \\ -2q & 1+(1-u^2)q^2
  \end{pmatrix}, \nn \\
  M_n &= \begin{pmatrix}
 1+(1-u^2)q^2  & -q    & \cdots & 0 & -q  \\
  -q     & 1+(1-u^2)q^2 & \cdots & 0 & 0 \\
  \vdots & \vdots& \ddots & \vdots & \vdots \\
  0     & 0      & \cdots & 1+(1-u^2)q^2   & -q \\
  -q     & 0     & \cdots & -q     & 1+(1-u^2)q^2
\end{pmatrix}\ (n\ge 3)\,.
\end{align}
It is straightforward to evaluate the determinant and the result is 
\begin{equation}
  \zeta_{C_n}(q,u) = \Bigl(\xi_+(q,u)^\frac{n}{2}-\xi_-(q,u)^\frac{n}{2}\Bigr)^{-2}\,,  
  \label{eq:zetaCn}
\end{equation}
where $\xi_\pm$ are the solutions of the quadratic equation $x^2-(1+(1-u^2)q^2)x + q^2 =0$; 
\begin{equation}
  \xi_\pm(q,u) \equiv \frac{1}{2}\Bigl( 1+(1-u^2)q^2
  \pm \sqrt{1-2(1+u^2)q^2 + (1-u^2)^2 q^4} \Bigr)\,.
\end{equation}

We next evaluate the partition function $Z_{\rm gKM}$ for $C_n$ by using the expression \eqref{eq:Z_zeta}. 
Since excluding the edge $e=1$ from $C_n$ yields the spanning graph of $C_n$, 
we can fix the gauge by setting $U_2=\cdots =U_n=1$
and the gauge field remains only on the first edge $e=1$.
We then denote $U_1=U$ in the following.
The matrix weighted Bartholdi zeta function $\zeta_{C_n}(q,u;U)$ in this gauge can be computed by using the same technique deriving \eqref{eq:zetaCn} as%
\footnote{In this computation, we have also used 
$\frac{1}{2q^{2n}}\left( 2q^n+\zeta_{C_n}(q,u)^{-1} -\sqrt{\zeta_{C_n}(q,u)^{-2}+ 4q^n \zeta_{C_n}(q,u)^{-1}}\right) = \xi_+(q,u)^{-n}\,.$}
\begin{align}
  \zeta_{C_n}(q,u;U) &= \det\Bigl(\zeta_{C_n}(q,u)^{-1} \bsone_{N^2}
  + (2-U\otimes U^\dagger - U^\dagger \otimes U)q^n\Bigr)^{-1}  \nn \\
  &= \xi_+(q,u)^{-N^2n}
  \Bigl| \det\Bigl(\bsone_{N^2} - q^n \xi_+(q,u)^{-n}\, U\otimes U^\dagger \Bigr) \Bigr|^{-2}\nn \\
  &=\xi_+(q,u)^{-N^2n}
  e^{2\sum_{k=1}^\infty \frac{1}{k}(q^n \xi_+(q,u)^{-n})^k |\Tr U^k|^2}\,, 
  \label{eq:zetaU_Cn}
\end{align}
where we have used $\zeta_{C_n}(q,u)$ is expressed as \eqref{eq:zetaCn}.
Substituting \eqref{eq:zetaU_Cn} to \eqref{eq:Z_zeta} we obtain
\begin{align}
  Z_{\rm gKM} &=
  \left(\frac{2\pi}{\beta}\right)^{\frac{N^2}{2}n}
  \xi_+(q,u)^{-\frac{N^2}{2}n}
  \prod_{i=1}^N \Bigl({1-q^{ni} \xi_+(q,u)^{-ni}}\Bigr)^{-1}\,, 
  \label{eq:ZgKM_Cn}
\end{align}
where we have used the formula obtained in \cite{matsuura2022kazakov}, 
\begin{align}
  \int dU \, e^{\sum_{k=1}^\infty \frac{q^k}{k}|\Tr U^k|^2 }
&=\prod_{i=1}^N \frac{1}{1-q^i}\,, \quad \text{for }|q|<1.
\label{eq:intU}
\end{align}
Note that it is easy to see $\xi_+(q,0)=1$,  which reproduces the result obtained in \cite{matsuura2022kazakov}.

\subsection{Derivation using the covering graphs}

We can derive \eqref{eq:ZgKM_Cn} also from the dual expression \eqref{eq:IZrep}. 
In this case, the Bartholdi zeta functions of the covering graphs play an important role instead of the unitary matrix integral.

For $G=C_n$, we can fix the gauge by setting $\sigma_{e=2}=\cdots=\sigma_{e=n}=1$ and we write $\sigma_1=\sigma$. 
Then the partition function \eqref{eq:IZrep} becomes a summation of Gaussian integrals, 
\begin{align}
  Z_{\rm gKM} &= \frac{1}{N!} \frac{(2\pi)^{\frac{N(N-1)}{2}n}}{\beta^{\frac{N^2}{2}n}q^{\frac{N(N-1)}{2}n}}
   \int \lrpar{ \prod_{v=1}^n\prod_{i=1}^N d\phi_{v,i} }
  \sum_{\sigma_\in S_N} \sgn(\sigma)
  e^{-\frac{1}{2} \phi_{v,i}(D_\sigma)_{vi;v'j}\phi_{v',j}}\,. 
  \label{eq:tmp1}
\end{align}
As mentioned below \eqref{eq:Dsigma}, $D_\sigma$ is the matrix that associated with the Bartholdi zeta function \eqref{eq:vertex Bartholdi} of a covering graph of $C_n$. 
In this case, the recombination of the edges occurs only at the edge $e=1$ by the permutation $\sigma$. 
The permutation is a product of cyclic permutations in general and the elements of $S_N$ can be classified by a conjugacy class labeled by the number and the lengths of the cyclic permutations, that is, a partition of $N$. 
Let us write a partition of $N$ as $\lambda=(l_1^{m_1},\cdots,l_r^{m_r})$ where $m_i$ is the number of the cyclic permutations of the length $l_i$. 
From the construction, a cyclic permutation of the length $l$ connects $l$ layers of $C_n$ in turn which yields a cyclic graph $C_{ln}$. 
Therefore the covering graph appeared by a permutation $\sigma$ in the conjugacy class labeled by $\lambda$ is 
$(C_{nl_1})^{m_1}\oplus \cdots \oplus (C_{nl_r})^{m_r}$  and we see 
\begin{equation}
  (\det D_\sigma)^{-1} = \prod_{i=1}^r \zeta_{C_{nl_i}}(q,u)^{m_i}\,.
\end{equation}
Recalling that there are $N!/z_\lambda$ elements in the conjugacy class labeled by $\lambda$ with 
\begin{equation}
  z_\lambda \equiv \prod_{i=1}^r m_i!\, l_i^{m_i}\,,
 \label{eq:zlambda}
\end{equation}
and the signature of the permutation is written as ${\rm sgn}\ \sigma = \prod_{i=1}^r (-1)^{(l_i-1)m_i}=(-1)^N \prod_{i=1}^r (-1)^{m_i}$, 
we can perform the Gaussian integral in \eqref{eq:tmp1} as 
\begin{align}
Z_{\rm gKM} &= 
\frac{1}{N!} 
\left(\frac{2\pi}{\beta}\right)^{\frac{N^2}{2}n}
\frac{(-1)^N}{q^{\frac{N(N-1)}{2}n}}
\sum_{\lambda\vdash N}
\prod_{i=1}^r 
\frac{N!}{z_\lambda}
\left(-\zeta_{C_{nl_i}}(q,u)^{1/2}\right)^{m_i} \nn \\
&=
\left(\frac{2\pi}{\beta}\right)^{\frac{N^2}{2}n}
\frac{(-1)^N}{q^{\frac{N(N-1)}{2}n}}
\sum_{\lambda\vdash N}
\prod_{i=1}^r 
\frac{1}{z_\lambda}
\left(\frac{1}{\xi_-(q,r)^\frac{nl_i}{2}-\xi_+(q,r)^\frac{nl_i}{2}}\right)^{m_i} \nn \\
&=
\left(\frac{2\pi}{\beta}\right)^{\frac{N^2}{2}n}
\frac{(-1)^N}{q^{\frac{N(N-1)}{2}n}\xi_-(q,u)^\frac{Nn}{2}}
\sum_{\lambda\vdash N}
\prod_{i=1}^r 
\frac{1}{z_\lambda}
\left(\frac{1}{1-
(\xi_+(q,u)^\frac{n}{2}/\xi_-(q,u)^\frac{n}{2})^{l_i}}\right)^{m_i}\,,
\label{eq:tmp2}
\end{align}
where we have used \eqref{eq:zetaCn} and $\xi_-\le\xi_+$ from the first line to the second line.
To evaluate the summation in \eqref{eq:tmp2}, we can use the identity, (see e.g. \cite{matsuura2022kazakov} for a proof)
\begin{equation}
  \sum_{\lambda\vdash N}
  \prod_{i=1}^r 
  \frac{1}{z_\lambda}
  \left(\frac{1}{1-x^{l_i}}\right)^{m_i} 
  = \prod_{i=1}^N \frac{1}{1-x^i}\,.
\end{equation}
Combining $\xi_-(q,u)\xi_+(q,u)=q^2$, 
we finally obtain 
\begin{align}
  Z_{\rm gKM} &=
  \left(\frac{2\pi}{\beta}\right)^{\frac{N^2}{2}n}
  \xi_+(q,u)^{-\frac{N^2}{2}n}
  \prod_{i=1}^N \Bigl({1-q^{ni} \xi_+(q,u)^{-ni}}\Bigr)^{-1}\,, \nn
\end{align}
for $C_n$, which is identical with \eqref{eq:ZgKM_Cn}.

\subsection{Counting the cycles and the Catalan numbers}

The expression \eqref{eq:ZgKM_Cn} is the realization of \eqref{eq:path W-Bartholdi} for $G=C_n$, namely ${\cal V}_{C_n}(q,u)=\xi_+(q,u)^{-n}$ and $f_C(q,u)=q^n \xi_+(q,u)^{-n}$. 
We can explicitly see that the function $\xi_+(q,u)^{-n}$ and $q^n \xi_+(q,u)^{-n}$ count such cycles that reduce to a vertex 
(we refer to them as ``collapsing cycles'' in the following)
and the primitive reduced cycle $C$ of $C_n$, respectively.

Let us first consider the case of $n=1$.
The point is that the function  
\begin{equation}
  \xi_+(q,u)^{-1} = \frac{\xi_-(q,u)}{q^2} 
  =  \frac{1+(1-u^2)q^2-\sqrt{1-2(1+u^2)q^2+(1-u^2)^2 q^4}}{2q^2}\,, 
  \label{eq:general Catalan}
\end{equation}
is the generating function of the generalized Catalan numbers \cite{barry2019generalized}. 
The usual Catalan number ${\rm Cat}(m)$ can be interpreted as the number of the Dyck words of length $2m$,
that is, such sequences of two letters $X$ and $Y$ of the same number $(m)$ such that no initial segment of the sequence has more $X$ than $Y$,  
which appears as the coefficient of $q^{2m}$ when expanding the expression \eqref{eq:general Catalan} by $q$ with $u=1$.
The generalized Catalan number also counts the number of the borders of $X$'s and $Y$'s in the Dyck words, which appears in the coefficient of $u$ in the expansion of \eqref{eq:general Catalan}. 
This is exactly the way to construct collapsing primitive cycles from the two primitive reduced cycles $C$ and $C^{-1}$ of $G=C_1$. 
Since the powers of $q$ and $u$ can be naturally interpreted as the length of the cycle and the number of the bumps, respectively, 
we can conclude that $\xi_+(q,u)^{-1}$ counts the collapsing cycles. 
The same is true with respect to the element of $[\cB_C]$
and $q\xi_+(q,u)^{-1}$ counts the cycles in $[{\cal B}_C]$, 
since any cycle $\tC\in [{\cal B}_C]$ is constructed by putting a collapsing primitive cycle to $C$.

We can generalize this argument to an arbitrary $n$: 
The collapsing cycles of $G=C_n$ can be again counted by Dyck words. 
To see it, we write the $n$ edges of $G=C_n$ as $e_i$ ($i=1,\cdots,n$). 
A path of $C_n$ can be expressed as a sequence of $e_i$, 
but there is a strict rule for the order of the sequence: 
After $e_i$, only $e_{i+1}$ or $e_i^{-1}$ can appear, and after $e_i^{-1}$, only $e_i$ or $e_{i-1}^{-1}$ can appear.
Therefore, if we fix the starting point of the path, 
we can omit the subscript $i$ of the edges to express the path. 
As a result, the path is uniquely specified as the starting point and a sequence of $e$ and $e^{-1}$. 
In particular, if the path is a collapsing cycle, it is expressed as a Dyck word by properly rotating the cycle if necessary. 
Since $C_n$ has $n$ vertices, we see ${\cal V}_{C^n}(q,u)=\xi_+(q,u)^{-n}$. 

We next consider a path $\tC\in [\cB_C]$ with $C=(e_1,\cdots,e_n)$. 
The key observation is that $\tC$ always contains the pairs $e_i e_{i+1}$ ($i=1,\cdots,n$). 
Therefore, by rotating the order of $e_i$'s properly, we can express $\tC$ as 
\begin{equation}
\tC = (w_1, e_1, w_2, e_2, \cdots ,w_n, e_n)\,,\nn
\end{equation}
where $w_i$ is a collapsing cycle starting from $e_i$ and not including the pair $e_i e_{i+1}$. 
As mentioned above, we can omit the indices of the edges to express the path. 
Using this rule, $\tC$ corresponds to a word made of $e$ and $e^{-1}$, 
\[
d_1(e,e^{-1}) e\, d_2(e,e^{-1}) e\, \cdots d_n(e,e^{-1})e\,,
\]
where $d_i(e,e^{-1})$ is a Dyck word made of $e$ and $e^{-1}$, which comes from $w_i$. 
Therefore, there is one-to-one correspondence between the element of $[\cB_C]$ and $n$ set of Dyck words, 
which yields $f_C(q,u)=q^n \xi_+(q,u)^{-n}$.

\subsection{The partition function in the parametrization of the original KM model}

Although the parametrization of the gKM model adopted in \eqref{eq:SgKM} is convenient to see the relation between the gKM model and the Bartholdi zeta function, 
it is worth looking at how the partition function appears in the parametrization of the original KM model, since the cycle graph $C_n$ we are considering is nothing but the one-dimensional square lattice
where the exact form of the partition function is known \cite{caselle1992exact}.
We then consider the action \eqref{eq:SgKM reguar} with $d=2$ and the mass parameter \eqref{eq:mass} becomes 
\begin{equation}
  m^2 = q^{-1} + q (1-u^2)\,. 
  \label{eq:muv}
\end{equation}


This reparametrization makes the expressions simpler. 
Since the generating function of the generalized Catalan numbers \eqref{eq:general Catalan} 
can be rewritten as 
\begin{equation}
  \xi_+(q,u)^{-1} = \frac{R(m^2)}{q}\,,
\end{equation}
where
\begin{equation}
  R(m^2) \equiv \frac{m^2-\sqrt{m^4-4}}{2} \,,
  \label{eq:funcR}
\end{equation}
the partition function \eqref{eq:ZgKM_Cn} can be written as 
\begin{align}
  Z_{\rm gKM}
  &= \left(\frac{2\pi}{\beta q}\right)^{\frac{N^2}{2}n}
  R(m^2)^{\frac{N^2}{2}n}
  \prod_{i=1}^N \Bigl({1-R(m^2)^{ni}}\Bigr)^{-1}\,,
  \label{eq:ZgKM_Cn-R}
\end{align}
for the cycle graph $C_n$,
which reproduces the result in \cite{caselle1992exact}.
We can explicitly see that the identities \eqref{eq:formulae1} (and \eqref{eq:formulae2}) hold.

We point out that the function $R(m^2)$ is the Stieltjes transformation of the semi-circle distribution (or the expectation value of the resolvent); 
\begin{equation}
  R(m^2) = \frac{1}{2\pi} \int_{-2}^2 dx \frac{\sqrt{4-x^2}}{m^2-x}  \,.
  \label{eq:resolvent}
\end{equation}
It is plausible to think that this is related to the result \cite{gross1992some} that the semi-circle distribution is the exact solution of the eigenvalue distribution of the scalar fields of the original KM model in large N (see Sect.~\ref{sec:largeN}).
It is interesting, however, that the function representing the semi-circle distribution already appears in the partition function at finite $N$, which is a different situation from the Gaussian matrix model. 
This reflects the fact that the one-dimensional KM model cannot be completely regarded as a Gaussian matrix model.
In fact, as already mentioned, we cannot fix all unitary matrices of a one-dimensional KM model but there always remains a unitary matrix on one edge. This is why the partition function \eqref{eq:ZgKM_Cn} or \eqref{eq:ZgKM_Cn-R} does not turn out to be that of the Gaussian matrix model.

Another insight obtained by this reparametrization is the relation between the Catalan numbers and the generalized Catalan numbers: 
Recalling that the Catalan numbers are obtained as the moments of the semi-circle distribution, 
\begin{equation}
  \frac{1}{2\pi}\int_{-2}^2 dx\, x^k \sqrt{4-x^2} = 
  \begin{cases}
    {\rm Cat}\left(\frac{k}{2}\right) & \text{$k$: even} \\
    0 & \text{$k$: odd} \\
  \end{cases}\,,
\end{equation}
we can rewrite the generating function \eqref{eq:general Catalan} as 
\begin{equation}
  \xi_+(q,u)^{-1} 
  = \frac{1}{q} \sum_{k=0}^\infty \frac{{\rm Cat}(k)}{(m^2)^{2k+1}}
  = \frac{1}{q} \sum_{k=0}^\infty \frac{{\rm Cat}(k)}{(q^{-1}+q(1-u^2))^{2k+1}}\,, 
\end{equation}
where we have used \eqref{eq:muv}.
This gives a direct relation between the Catalan numbers and the generalized Catalan numbers. 

\section{Graph Kazakov-Migdal model at large $N$}
\label{sec:largeN}

\subsection{Exact evaluation of the partition function}

We next evaluate the partition function \eqref{eq:ZgKM} for an arbitrary graph $G$ in the limit of $N\to\infty$ by carrying out the unitary matrix integral of \eqref{eq:Z_zeta}.
To this end, we denote the Wilson loop along a primitive reduced cycle $C$ of the graph as $P_C(U)$,
and introduce the notations,
\[
  |\Tr f(U) |^2 \equiv \Tr f(U) \,\Tr f(U)^{\dagger}\,,
\]
and
\[
  \Bigl\langle
   g(U)
  \Bigr\rangle \equiv \int \prod_{e\in E} dU_e\, g(U)\,,
\]
where $f(U)=f(U_1,\cdots,U_{n_E})$ and
$g(U)=g(U_1,\cdots,U_{n_E})$ are functions of $U_e$ ($e=1,\cdots,n_E$).

Although it seems impossible at first glance to perform the complex integral involving multiple unitary matrices,
a significant simplification occurs at large $N$ \cite{Kazakov:1983fn, KOSTOV1984191, OBRIEN1985621}.
The essential facts are that the integral of $|\Tr P_C(U)^{l}|^2$
at large $N$ as
\begin{equation}
  \Bigl\langle
    |\Tr P_C(U)^{l} |^2
  \Bigr\rangle 
  \underset{N\to\infty} \longrightarrow l \,,
  \label{eq:PCl}
\end{equation}
and
that the integral of the product of the two Wilson loops along two (not necessarily primitive) cycles $C_1$ and $C_2$ are decomposed into the product of
the integrals of the individual Wilson loops
at large $N$ as
\begin{align}
 \Bigl\langle
   \bigl|\Tr P_{C_1}(U)
   \Tr P_{C_2}(U) \bigr|^2
 \Bigr\rangle
 \underset{N\to\infty}{\longrightarrow}
 \Bigl\langle
   \bigl|\Tr P_{C_1}(U)\bigr|^2
 \Bigr\rangle
 \Bigl\langle
   \bigl|\Tr P_{C_2}(U)\bigr|^2
 \Bigr\rangle\,.
 \label{eq:factorization}
\end{align}
In general, we can define a general Wilson loop along a primitive cycle $C$ associated with a partition
$\lambda=(l_1^{m_1} l_2^{m_2}\cdots)$ by
\[
  \Upsilon_{\lambda}(P_C(U)) \equiv \prod_{i}
  (\Tr P_C(U)^{l_i})^{m_i}\,.
\]
Then, combining \eqref{eq:PCl} and \eqref{eq:factorization},
we can evaluate the integral of a product of
general Wilson loops of the chiral primitive cycles at large $N$ as
\begin{align}
\Bigl\langle \prod_{C\in \Pi^+}| \Upsilon_{\lambda}(P_C(U))|^2 \Bigr\rangle
  \underset{N\to\infty} \longrightarrow
  \prod_{C\in \Pi^+} z_{\lambda}\,,
  \label{eq:asifDS}
\end{align}
where $z_\lambda$ is defined by \eqref{eq:zlambda}. 
See \cite{matsuura2022kazakov} for more detail.

We can evaluate \eqref{eq:Z_zeta} by using these facts.
Since the matrix weighted Bartholdi zeta function can be written as \eqref{eq:path W-Bartholdi}, we can evaluated the integration in \eqref{eq:Z_zeta} as
\begin{align}
  {\cal V}_{G}(q,u)^{-\frac{N^2}{2}} &\int \prod_{e\in E} dU_e\, \zeta_G(q,u;U)^\frac{1}{2} \nn \\
  &=
  \Bigl\langle
    \prod_{C\in\Pi^+} e^{\sum_{m=1}^\infty \frac{1}{m}f_C(q,u)^m |\Tr P_C(U)^m|^2}
  \Bigr\rangle \nn \\
  &=
  \Bigl\langle
    \prod_{C\in\Pi^+} \sum_{n=1}^\infty 
    f_C(q,u)^n
    \sum_{\lambda\vdash n}
    \frac{1}{z_\lambda}|\Upsilon_\lambda(P_C(U))|^2
  \Bigr\rangle \nn \\
  &\underset{N\to\infty} \longrightarrow
    \prod_{C\in\Pi^+} \sum_{n=1}^\infty 
    {f_C(q,u)^n}p(n)
  =
  \prod_{C\in\Pi^+} 
  \prod_{k=1}^\infty 
  \frac{1}{1-f_C(q,u)^k}\,,
  \label{eq:tochu3}
\end{align}
where 
we have use the large $N$ decomposition \eqref{eq:asifDS} in the final line
and $p(n)$ is the number of the partitions of $n$.
Therefore, the partition function \eqref{eq:ZgKM} in the large $N$ limit behaves as
\begin{align}
  Z_{\rm gKM}
   \underset{N\to\infty} \longrightarrow 
   \left(\frac{2\pi}{\beta}\right)^{\frac{1}{2}n_V N^2}
    (1-(1-u)^2q^2)^{\frac{1}{2}{(n_E-n_V)N^2}}
  {\cal V}_{G}(q,u)^{\frac{N^2}{2}}
  \prod_{C\in\Pi^+} \prod_{k=1}^\infty  \frac{1}{1-f_C(q,u)^k}\,.
  \label{eq:ZgKM_largeN}
\end{align}
Note that, the partition function of the gKM model on a $d$-regular graph \eqref{eq:ZgKM_reular} can be evaluated using the same technique as 
\begin{align}
  Z_{\rm gKM} &\underset{N\to\infty}{\longrightarrow}
  \left(\frac{2\pi}{\beta q\, m^2}\right)^{\frac{1}{2}n_V N^2}
  {\cal V}_G(m^{-2},1)^{\frac{N^2}{2}}
  \prod_{C\in\Pi_+} \prod_{k=1}^\infty  \frac{1}{1-f_C(m^{-2},1)}\,. 
  \label{eq:ZgKM_reular_largeN}
\end{align}

\subsection{Regularized partition function}

Since there is still $N$-dependence on the right-hand side of \eqref{eq:ZgKM_largeN}, 
the precise meaning of this expression is that the partition function of the gKM model is asymptotic to the right-hand side at large $N$.
Therefore, it is convenient to give a suitable approximation of the partition function at finite $N$ which asymptotically
approach the right-hand side of \eqref{eq:ZgKM_largeN}. 

The part 
to consider is 
$\prod_{C\in\Pi^+} \prod_{k=1}^\infty  \frac{1}{1-f_C(q,u)^k}$ 
in the right-hand side, which is obtained by evaluating 
\begin{equation}
  \left\langle 
  \prod_{C\in[{\cal P}_R]} \sum_{n=0}^\infty f_C(q,u)^n \sum_{\lambda \vdash n} \frac{1}{z_\lambda} | \Upsilon_\lambda( P_C(U) )|^2
  \right\rangle\,.
  \label{eq:tochu6}
\end{equation}
We can safely use the large $N$ decompositions \eqref{eq:PCl} and \eqref{eq:factorization} because all the class functions $\Upsilon_\lambda( U )$ for $U\in U(N)$ are independent regardless of the size of the partition $\lambda$.  
However, at finite $N$, we have to take care of the size of the partition 
since $\Tr(U^n)$ for $n>N$ can be expanded by products of $\Tr(U^k)$'s ($k\le N$)%
\footnote{
This is because $\Tr(U^n)$ is the $n$'s power-sum symmetric polynomial of the $N$ eigenvalues of $U$.}. 
We therefore restrict the partitions to those with row lengths less than or equal to $N$ so that we count only independent Wilson loops. 
If we ignore the ${\cal O}(1/N)$ contributions, we can further use the large $N$ decomposition \eqref{eq:asifDS} as an approximation. 
Then we can approximate \eqref{eq:tochu6} as 
\begin{align}
  \left\langle 
  \prod_{C\in[{\cal P}_R]} \sum_{n=0}^\infty f_C(q,u)^n \sum_{\lambda \vdash n} \frac{1}{z_\lambda} | \Upsilon_\lambda( P_C(U) )|^2
  \right\rangle
  &\simeq 
  \prod_{C\in[{\cal P}_R]} \sum_{n=0}^\infty f_C(q,u)^n 
  \sum_{\underset{|\lambda_i|\le N}{\lambda \vdash n} }
    \frac{1}{z_\lambda} 
  \left\langle 
| \Upsilon_\lambda( P_C(U) )|^2
  \right\rangle \nn \\
  &= 
  \prod_{C\in[{\cal P}_R]} \sum_{n=0}^\infty f_C(q,u)^n 
  p(n;N) \nn \\
  &=\prod_{C\in[{\cal P}_R]} \prod_{k=1}^N \frac{1}{1-f_C(q,u)^k}\,, 
\end{align}
where $p(n;N)$ is the number of Young tableau with $n$ boxes and the lengths of the rows are less than or equal to $N$.
We then propose the approximate partition function at finite $N$ as 
\begin{align}
  \hat{Z}_{\rm gKM}
   \equiv 
   \left(\frac{2\pi}{\beta}\right)^{\frac{1}{2}n_V N^2}
    (1-(1-u)^2q^2)^{\frac{1}{2}{(n_E-n_V)N^2}}
  {\cal V}_{G}(q,u)^{\frac{N^2}{2}}
  \prod_{C\in\Pi^+} \prod_{k=1}^N  \frac{1}{1-f_C(q,u)^k}\,. 
  \label{eq:ZgKM_N}
\end{align}
It is obvious that \eqref{eq:ZgKM_N} asymptotically approaches the right-hand side of \eqref{eq:ZgKM_largeN}%
\footnote{
The difference between the regularized free energy \eqref{eq:free energy general} and the exact free energy at finite $N$ is ${\cal O}(1)$ because the large $N$ decomposition is justified by ignoring ${\cal O}(1/N)$ contributions from expectation values. Therefore we can use the regularized partition function \eqref{eq:ZgKM_N} if we are interested in the leading and the next leading behavior in $1/N$-expansion. }
in the large $N$ limit. 
Furthermore, the exact partition function of the model for $G=C_n$ at finite $N$ coincide to \eqref{eq:ZgKM_N}. 
It supports that the expression \eqref{eq:ZgKM_N} can be regarded as a regularized partition function of the gKM model at finite $N$.

The free energy calculated from this regularized partition function becomes
\begin{align}
  \hat{\cal F}_{\rm gKM} &\equiv -\frac{2}{n_V} \log \hat{Z}_{\rm gKM} \nn \\
  &\simeq 
  N^2\left(-\frac{1}{n_V}\log{\cal V}_G(q,u)\right)
  + \frac{2}{n_V} \sum_{C\in \Pi_+}\sum_{k=1}^N \log\left(1-f_C(q,u)^k\right)\,, 
  \label{eq:free energy general}
\end{align}
up to irrelevant terms. 
It is remarkable that the leading ${\cal O}(N^2)$ contribution to the free energy comes from ${\cal V}_G(q,u)$ which counts the collapsing cycles, 
which explicitly realizes the analysis given in \cite{boulatov1993infinite,balakrishna1994difficulties}.
On the contrary, the second term is the contribution from non-zero area Wilson loops but is ${\cal O}(N)$. 
This suggests that, if we deal only with the leading of the 1/N expansion of this theory, we completely ignore the physics arising from the non-zero area Wilson loop. We will see it explicitly in the next subsection.

\subsection{Relation to the exact solution at large $N$}

For the original KM model \eqref{eq:SKM} defined on the $D$-dimensional square lattice, 
the equations satisfied by the density of eigenvalues of the scalar field for large $N$ are derived in  \cite{migdal1993exact}, 
and it was found that the semi-circle distribution satisfies the equations for any $D$ in \cite{gross1992some}.
Let us first briefly review this analysis.

The basic idea is to combine the equations satisfied by the IZ integral, 
\[
I(\Phi,\Psi) \equiv \int dU e^{N\Tr \left(\Phi U \Psi U^\dagger \right)}\,, 
\]
with the saddle point equation satisfied by the KM model at large $N$.
More concretely, the equations are those for the matrix functions, 
\begin{align}
  F(\Phi) \equiv \nabla_\Phi \ln I(\Phi,\Psi), \quad 
  G_\lambda(\Phi) \equiv \frac{1}{\lambda-F(\Phi)-\nabla_\Phi }\cdot 1 \,, \nn
\end{align}
with $(\nabla_\Phi)_{ij}\equiv \frac{1}{N}\frac{\partial}{\partial \Phi_{ji}}$. 
We introducing the density function of the eigenvalues of $\Phi$, 
\begin{equation}
\rho(\mu) \equiv \frac{1}{N} \sum_{a=1}^N \delta(\mu-\phi_a)\,.  \nn
\end{equation}
Then, assuming that all $\Phi(x)$ have the same eigenvalue distribution in the large $N$ limit, 
we can show that the IZ integral satisfies the following equations at large $N$: 
\begin{align}
  {\rm Re}V'(z) &= {\cal P}\int_{-\infty}^\infty \frac{d\nu}{2\pi i} \log\left(
  \frac{z-F(\nu)-{\rm Re}V'(\nu)+i\pi \rho(\nu)}
  {z-F(\nu)-{\rm Re}V'(\nu)-i\pi \rho(\nu)}
  \right), \quad 
  {\rm Im}V'(z) = -\pi \rho(z)\,,
  \label{eq:eqIZ}
\end{align}
where $V'(\lambda) \equiv \frac{\rho(\nu)}{\lambda-\nu}$ and ${\cal P}\int$ denotes the principal integration. 
Furthermore, from the action \eqref{eq:SKM}, we obtain the saddle point equation in the large $N$, 
\begin{equation}
  F(z) = \frac{1}{D}\left(\frac{1}{2} U'(z) - {\rm Re} V'(z)\right)
  = \frac{1}{D}\left(m^2 z - {\rm Re} V'(z)\right)
  \,, 
  \label{eq:original saddle point}
\end{equation}
where $U(\Phi)$ is the potential of the KM model and is quadratic, $U(\Phi)=\frac{m^2}{2} \Phi^2$, in the present case
and the second term ${\rm Re}V'(z)$ comes from the variation of the Vandermonde determinant. 
Substituting \eqref{eq:original saddle point} to \eqref{eq:eqIZ}, we can eliminate the dependence of $F$ from the equations. 
For more detail, see \cite{migdal1993exact}. 

In \cite{gross1992some}, it has been shown that the semi-circle distribution,
\begin{equation}
  \rho(\nu) = \frac{1}{\pi}\sqrt{\mu-\frac{\mu^2\nu^2}{4}}\,,
  \label{eq:semi-circle}
\end{equation}
solves the equations \eqref{eq:eqIZ} and \eqref{eq:original saddle point} for any $D$ as 
\begin{equation}
  \mu_\pm = \frac{1}{2D-1}\Bigl(
  (D-1) m^2 \pm D\sqrt{m^4-8D+4}
  \Bigr)\,,
\end{equation}
where $\mu$ is a constant determined by $D$ and $m$.

We can apply the same analysis to the gKM model. 
To justify the assumption that all the scalar field have the same eigenvalue distribution, 
we consider a $d$-regular graph as in Sec.~\ref{sec:regular}. 
Then the saddle point equation of the gKM model becomes 
\begin{equation}
  F(z) = \frac{2}{d}\left(
  m^2 z - {\rm Re}V'(z)
  \right)\,. \nn
\end{equation}
Comparing it to the saddle point equation for the original KM model \eqref{eq:original saddle point}, 
we see that the equations can be solved by simply replacing $D$ with $d/2$.
Therefore, we can conclude that 
the semi-circle distribution \eqref{eq:semi-circle}
with 
\begin{equation}
  \mu = \mu_\pm = \frac{1}{2(d-1)}\Bigl(
  (d-2) m^2 \pm d\sqrt{m^4-4(d-1)}
  \Bigr)\,,
\end{equation}
is an exact solution of the gKM model on a $d$-regular graph at large $N$.
This indicates that the semi-circle distribution is a universal scalar field behavior independent of the details of the graph.

What type of Wilson loops are involved in this semi-circle solution?
As discussed in \cite{gross1992some}, the contribution to free energy from the semi-circle solution \eqref{eq:semi-circle} is 
\begin{align}
  {\cal F}_{\rm SC}=N^2\left(
  \frac{1}{2}\frac{m^2}{2\mu}+\frac{1}{2}\log\mu -\frac{D}{2}\sqrt{1+\frac{4}{\mu^2}}-1-\log\left(\frac{1}{2}+\frac{1}{2}\sqrt{1+\frac{4}{\mu^2}}\right)
  \right)\,, \nn
\end{align}
which is ${\cal O}(N^2)$. 
The $1/N$-expansion of the free energy \eqref{eq:free energy general} suggests that the semi-circle solution comes not from the non-zero area Wilson loops but from the zero area Wilson loops. 
To confirm it, let us evaluate the expectation value of the scalar fields, 
\begin{equation}
  \frac{1}{N n_V}\Bigl\langle
   \sum_{v\in V} \Tr\Phi_v^2  
  \Bigr\rangle
  =
  \frac{1}{N n_V}\Bigl\langle
     \sum_{v\in V}\sum_{a=1}^N \phi_{v,a}^2
  \Bigr\rangle\,. 
\end{equation}
If the distribution of the eigenvalues is semi-circle \eqref{eq:semi-circle}, the expectation value can be evaluated as 
\begin{equation}
  \MM_{\rm SC}  = \frac{1}{\pi} \int_{-\frac{2}{\sqrt{\mu}}}^{\frac{2}{\sqrt{\mu}}} dx x^2 \sqrt{\mu-\frac{\mu^2 x^2}{4}} = \frac{1}{\mu}\,. \nn
\end{equation}
On the other hand, from the action \eqref{eq:SgKM reguar}, the expectation value can be evaluated as
\begin{align}
  \MM_{\rm gKM} 
  &= -\frac{2}{N^2 n_V} \frac{\partial}{\partial m^2} \log Z_{\rm gKM}  \nn \\
  &= -\frac{1}{n_V} \frac{\partial}{\partial m^2} \log {\cal V}_G(m^{-2},1) 
  +\frac{1}{N^2n_V} \sum_{C\in\Pi_+}\sum_{k=1}^N 
  \frac{\partial}{\partial m^2} 
  \log\left(1-f_C(m^{-2},1)^k\right)\,,
  \label{eq:tochu4}
\end{align}
where we have used the regularized partition function and set $\beta q=N$.
Since the functions ${\cal V}_G(m^{-2},1)$ and $f_C(m^{-2},1)$ depend on the details of the graph, 
we evaluate the expectation value for the cycle graph $C_n$ for which these functions can be explicitly written down as 
\begin{align}
  V_{C_n}(m^{-2},1)=R(m^2)^{\frac{N^2}{2}n}, \quad 
  f_{C}(m^{-2},1)=R(m^2)^n\,, \nn
\end{align}
where $R(m^2)$ is given by \eqref{eq:funcR}.
Substituting them into \eqref{eq:tochu4}, we obtain
\begin{align}
  \MM_{\rm gKM} 
  &= -\frac{R'(m^2)}{R(m^2)} \left( 
  1 - \frac{1}{N^2} \sum_{k=1}^N  \frac{R(m^2)^{nk}}{1-R(m^2)^{nk}} \right) 
  = \frac{1}{\mu} \left(1  + {\cal O}(1/N) \right)\,, 
  \label{eq:M2_Cn}
\end{align}
where we have used $\frac{-R'(m^2)}{R(m^2)} = \frac{1}{\sqrt{m^4-4}}=\frac{1}{\mu}$. This shows that the exact solution at large $N$ reflects only the zero area Wilson loops of the graph as expected. 

\subsection{Reduction to the Ihara zeta function}

As we have seen in Sect.~\ref{sec:Bartholdi zeta}, the Bartholdi zeta function reduces to the Ihara zeta function by setting $u=0$, which counts only the reduced cycles of the graph. In particular, the functions ${\cal V}_G(q,u)$ and $f_C(q,u)$ become ${\cal V}_G(q,0)=1$ and $f_C(q,0)=q^{|C|}$, respectively, and the (regularized) partition function \eqref{eq:ZgKM_N} becomes 
\begin{align}
    Z_{\rm gKM}^{G,N}
   &= 
   \left(\frac{2\pi}{\beta}\right)^{\frac{1}{2}n_V N^2}
    (1-q^2)^{\frac{1}{2}{(n_E-n_V)N^2}}
  \prod_{C\in\Pi^+} \prod_{k=1}^N  \frac{1}{1-q^{|C|k}} \nn \\
  &= 
     \left(\frac{2\pi}{\beta}\right)^{\frac{1}{2}n_V N^2}
    (1-q^2)^{\frac{1}{2}{(n_E-n_V)N^2}} \prod_{k=1}^N \zeta_G(q^k)^{\frac{1}{2}}\,,
\end{align}
as shown in \cite{matsuura2022kazakov}. 
We saw in the previous subsection that the semi-circle distribution is an exact solution of this model at large $N$ and it arises from the collapsing cycles.
A natural question then arises as to what happens to this solution when the collapsing cycles vanish at $u=0$. 

If the value of $u$ is fixed, 
we cannot evaluate $\MM$ simply by the derivative of the mass parameter, 
since the coefficients of $\sum_{v\in V}\Tr\Phi_v^2$ and $\sum_{e\in E} \Tr \Phi_{s(e)} U_e \Phi_{t(e)} U_e^{-1}$ are determined only by the single parameter $q$.
Instead, we can use the identity satisfied by a $d$-regular graph, 
\begin{equation}
  \left(q\frac{\partial}{\partial q}-\beta\frac{\partial}{\partial\beta}\right)\left( \frac{1}{N^2} \log Z_{\rm gKM} \right)
  = \frac{1}{2}\left(q^{-1}-((1-u)d-(1-u)^2)q)\right) \frac{\beta q}{N^2} \Bigl\langle \sum_{v\in V} \Tr\Phi_v^2 \Bigr\rangle\,.
  \label{eq:tochu5}
\end{equation}
In order to compare to the result \eqref{eq:M2_Cn} in the previous subsection, we consider the case of $u=0$ for $d=2$. 
In this parametrization, $\log Z_{\rm gKM}$ is written as 
\[
\frac{1}{N^2} \log Z_{\rm gKM}^{C_n} = \frac{n}{2}  \log\frac{2\pi}{\beta} + {\cal O}(1/N)\,. 
\]
where the ${\cal O}(1/N)$ part is independent of $\beta$. 
Since the left-hand side of \eqref{eq:tochu5} becomes
\[
q\frac{\partial}{\partial q}\left( \frac{1}{N^2} \log Z_{\rm gKM}^{C_n} \right) = {\cal O}(1/N), \quad
-\beta \frac{\partial}{\partial \beta}\left( \frac{1}{N^2} \log Z_{\rm gKM}^{C_n} \right) = \frac{n}{2} \,.
\]
we can evaluate the expectation value of the scalar fields as 
\begin{equation}
  \MM_{\rm gKM} = \frac{1}{q^{-1}-q} + {\cal O}(1/N) = \frac{1}{\mu} + {\cal O}(1/N)\,,
\end{equation}
where we have set $\beta q=N$ after the computation and used $\mu=\sqrt{m^4-4}=q^{-1}-q$ ($|q|<1$) for $u=0$.
Therefore, although we have seen only the second momentum of the eigenvalues, it is natural to think that the semi-circle distribution is still a solution of this model even if $u=0$.

However, this solution is of course nothing to do with the Wilson loops since the Ihara zeta function does not count the collapsing cycles; $V_G(q,u=0)=1$.
The reason why the semi-circle is still a solution for $u=0$ is because the leading part of the free energy includes the contribution of the Gaussian integral.
In fact, the radius of the semi-circle distribution obtained for $d=2$ is the same as the radius of the semi-circle distribution produced in the large $N$ limit of the Gaussian model, obtained by simply neglecting the contribution of the link variable from the action of the gKM model.
When $d$ is greater than 2, the semi-circle is modified by the additional term proportional to $(n_E-n_V)\log(1-(1-u)^2q^2)$ from that of the the Gaussian model. 
However, it is still true that no Wilson loop contributes to the solution for $u=0$.

\section{Conclusion and Discussion}
\label{sec:Conclusion and Discussion}

In this paper, we proposed an extension of the Bartholdi zeta function by putting matrices as weights on the edges of the graph and gave its determinant representation.
We showed that the partition function of the Kazakov-Migdal model defined on a graph (gKM model) is represented by the unitary matrix integral of the matrix weighted Bartholdi zeta function in general.
We also derived the dual expression of the partition function by integrating the unitary matrices first using the so-called HCIZ integration formula. 
We computed the partition function of the gKM model for the cycle graph $C_n$ exactly at finite $N$, and showed that the partition function is closely related to the generalized Catalan numbers.
As a byproduct, we found a relation between the generating function of the generalized Catalan numbers and that of the usual Catalan numbers.
We also evaluated the partition function of the gKM model for arbitrary graphs at large $N$ and showed that it can be expressed as an infinite product of deformed Ihara zeta functions.
We showed explicitly that the leading terms in the $1/N$-expansion of the free energy are the contribution of zero area Wilson loop, 
while the contribution of non-zero area Wilson loops is of the order of ${\cal O}(1/N)$ compared to the leading. 
We applied the large $N$ analysis performed on the original KM model to the gKM model and found that the semi-circle distribution is an exact solution for the gKM model on the regular graphs in general.
The contribution from the zero area Wilson loops disappears if we set $u=0$, where the gKM model is expressed by the Ihara zeta function.

The partition function of the gKM model is represented by the unitary matrix integral \eqref{eq:Z_zeta}, while at the same time having the dual expression \eqref{eq:IZrep}.
It is not surprising that the dual expression \eqref{eq:IZrep} contains the sum of the symmetry groups 
because unitary matrix integrals can be written in terms of symmetry groups in general as a result of the Schur-Weyl duality. 
However, it is interesting that the resulting expression includes the adjacency matrix \eqref{eq:Dsigma} of the covering graph.
This strongly suggests that the Bartholdi zeta function weighted by the unitary matrix is closely related to the Bartholdi zeta function of the covering graph.
As mentioned below \eqref{eq:Dsigma}, a covering graph is generated by assigning a permutation on each edge of the graph $G$. 
More generally, 
we can assume that an element of the finite group $\Gamma$ is assigned on the edge $e$ of the graph as $\alpha(e)\in \Gamma$.
When $\alpha(e^{-1})=\alpha(e)^{-1}$, this is called the ordinary voltage assignment in graph theory. 
Since any finite group can be represented by a permutation group, we can easily see that the ordinary voltage assignment $\alpha$ induces a covering graph $G^\alpha$ in the same way. 
The Bartholdi zeta function of $G^\alpha$ is closely related to the base graph $G$ as expected. 
It is achieved by defining the Bartholdi $L$-function \cite{sato2006weighted}, 
\begin{equation}
    \zeta_G(q,u,\alpha,\rho) \equiv \prod_{C\in [{\cal P}]} \det\left(\bsone - q^{|C|} u^{b(C)} \rho(\alpha(C)) \right)^{-1}\,,
    \label{eq:L-func}
\end{equation}
where $\alpha(C)$ is defined as $\alpha(C)\equiv \alpha(\bse_1)\cdots\alpha(\bse_r)\in\Gamma$ when the cycle is written as $C=\bse_1\cdots \bse_r$. 
As shown in \cite{sato2006weighted}, 
the Bartholdi zeta function of $G^\alpha$ can be expressed through the $L$-function as 
\[
\zeta_{G^\alpha}(q,u) = \prod_\rho \zeta_{G}(q,u,\alpha,\rho)^{\dim \rho}\,,
\]
where $\dim\rho$ is the dimension of the representation $\rho$. 
The similarity between the matrix weighted Bartholdi zeta function \eqref{eq:W-Bartholdi} and the Bartholdi $L$-function \eqref{eq:L-func} is obvious: 
The matrix weighted Bartholdi zeta function would be regarded as an $L$-function when the voltage assignment to the graph is extended to a Lie group. 
It is plausible to think that the appearance of the adjacency matrix \eqref{eq:Dsigma} of the covering graph will be a consequence of this similarity. 
A more comprehensive understanding of the graph zeta function from this perspective would be a promising topic for the future.

In this paper, we have restricted the parameters of the gKM model as \eqref{eq:SgKM} in order to reveal the relation between the gKM model and the Bartholdi zeta function, but it is not a general parametrization. 
The most general parametrization is to assign an independent coupling constant to all vertices and all edges as 
\begin{align}
  S_{gKM} = \Tr\left\{ \frac{1}{2}\sum_{v\in V} m_v^2 \Phi_v^2 - \sum_{e\in E} q_e \Phi_{s(e)} U_e \Phi_{t(e)} U_e^{-1}  \right\}\,.
  \label{eq:general action}
\end{align}
Using these degrees of freedom, we will be able to see a connection to a more general graph zeta function than the Bartholdi zeta function.
For example, the Bartholdi zeta function is generalized in \cite{sato2016generalized} where a different value of the bump parameter $u_v$ is assigned to every vertex $v\in V$ as 
\begin{equation}
  \zeta_{G}(q,u_1,\cdots,u_{n_V}) \equiv \prod_{C\in{\cal P}} \left( 1 - q^{|C|} \prod_{v\in V} u_v^{b_v(C)} \right)^{-1}\,,
\end{equation}
where $b_v(C)$ is the number of bumps $e_i=e_{i+1}^{-1}$ with the condition $t(e_i)=v$. 
We can reproduce (the matrix weighted version of) this zeta function by tuning the parameters of \eqref{eq:general action} as 
\begin{equation}
  m_v^2 = 1+q^2 \sum_{v':\text{neighbor of } v} \frac{1-u_{v'}}{1-q^2 (1-u_{v})(1-u_{v'})}, \quad
  q_e = \frac{q}{1-q^2 (1-u_{s(e)})(1-u_{t(e)})}\,,
\end{equation}
as the partition function up to an overall factor. 
It would be one of the future problems to examine the nature of the gKM model and compare it to the generalized Bartholdi zeta function.
The more interesting question, however, is for what category of parameters the partition function of the gKM model has an Euler product representation. Although the zeta function is generally defined through an Euler product, it is not obvious whether it has a determinant representation. Conversely, it is clear that the partition function of the gKM model can be expressed in determinant form, but it is not obvious whether it has a meaning as a zeta function or not. The Ihara and Bartholdi zeta functions are interesting zeta functions that have both Euler product and determinant representations, and the fact that they are associated with the gKM model suggests the possibility of shedding light on more general properties of the zeta function through this model. It is quite an interesting future problem to pursue this possibility.

We finally would like to make a comment on another possibility of the gKM model. 
The main part of the $1/N$ expansion of the gKM model reflects only zero area Wilson loops and the contribution of non-zero area Wilson loops, which is of most interest for gauge theory, appears rather in the sub-leading part of the $1/N$ expansion.
If we adjust the theory so that the contribution of the leading part is appropriately eliminated, 
we will be able to construct a sort of gauge theory in the continuum limit. 
It is of course not the usual QCD because of the existence of the center gauge symmetry, 
but it is still a summation of non-zero area Wilson loops.
In that case, the existence of the parameter $u$ is expected to cause interesting phenomena: 
From the Wilson loop's point of view, 
allowing bumps corresponds to considering the zigzag symmetry of the Wilson loop,
which is essential for gauge theories to behave like string theory in the large $N$ limit \cite{polyakov1998string}. 
It would be an interesting attempt in the future to use the graph zeta function to gain insight into the holographic principle.

\section*{Acknowledgments}
The authors would like to thank 
S.~Aoki,
S.~Koyama,
M.~Fukuma,
Y.~Furuta,
T.~Misumi,
S.~Nishigaki,
H.~Watanabe,
Y.~Yoshida
and
J.~Yumoto 
for useful discussions.
This work is supported in part
by Grant-in-Aid for Scientific Research (KAKENHI) (C), Grant Number 20K03934 (S.~M.).

\appendix
\section{A proof of Amitsur's theorem}
\label{app:amitsur}

In this appendix, we give a proof of Amitsur's theorem in order to make the paper self-contained. 

Let us consider words generated by the ``letters'' $\{1,\cdots,k\}$, 
that is, sequences of the $n$ letters. 
We call the number of the letters in a word the length of the word, 
and call a word primitive (or indecomposable) if it cannot be written as $w^n$ $(n\ge 2)$ 
for a word $w$ with a shorter length. 
We call the words $w_1$ and $w_2$ are equivalent when $w_1$ is obtained by a cyclic rotation of the letters in $w_2$. 
In the following, We denote the set of primitive words of length $d$ as $P_d$, 
the set of the representative of the primitive words of length $d$ as $L_d$, 
and the set of the representative of all primitive words as $L\equiv \bigcup_{d=1}^\infty L_d$.

Let us consider $n$ matrices $\{X_1,\cdots,X_k\}$. 
Corresponding to a word $w=i_1\cdots i_n$, we define 
\[
X_w \equiv X_{i_1}\cdots X_{i_n}\,. 
\]
Then, 
it is obvious that the quantity $(X_1+\cdots+X_k)^m$ is a summation of $X_w$ where $w$ runs all words of length $m$. 
Recalling that any word can be expressed as a positive power of a primitive word, 
we can write this quantity as 
\begin{equation}
(X_1+\cdots+X_k)^m = \sum_{d|m} \sum_{\tilde{w}\in P_d} X_{\tilde{w}}^{m/d}\,.
\label{eq:tochu}
\end{equation}
We here note that, if $w$ is a primitive word of length $d$, 
there are $d$ different equivalent words to  $w^n$ ($n\ge 1$).
Therefore, we can write the trace of the expression \eqref{eq:tochu} as
\begin{equation}
    \Tr (X_1+\cdots+X_k)^m = \sum_{d|m} d \sum_{w\in L_d} \Tr(X_w^{m/d})\,. \nn
\end{equation}
Using this result, we can show the identity, 
\begin{align}
  \log\det\left(\bsone-X_1-\cdots -X_k\right)^{-1}
  &= \sum_{m=1}^\infty\frac{1}{m} \Tr\left(X_1+\cdots +X_k\right)^m \nn \\
  &= \sum_{m=1}^\infty\frac{1}{m} \sum_{d|m} d \sum_{w\in L_d} \Tr(X_w^{m/d}) \nn \\
  &= \sum_{k=1}^\infty\frac{1}{k} \sum_{d=1}^\infty \sum_{w\in L_d} \Tr(X_w^k) \nn \\
  &= \log \prod_{w\in L} \det\left(\bsone-X_w\right)^{-1}\,. \nn
\end{align}
We then obtain 
\begin{equation}
    \det\left(\bsone-X_1-\cdots -X_k\right) =  \prod_{w\in L} \det\left(\bsone-X_w\right)\,. \nn 
\end{equation}

\section{Proof of the identity \eqref{eq:id fC}}
\label{app:fC}
In this appendix, we show the identity \eqref{eq:id fC}, that is, 
\begin{equation}
  f_{C,n}(q,u) = f_C(q,u)^n\, 
  \label{eq:resultfC}
\end{equation}
where 
\begin{equation}
\begin{split}
  f_{C,n}(q,u) &\equiv \sum_{k|n} k \sum_{\tC\in [\cB(C^k)]} (q^{|\tC|}u^{b(\tC)})^{n/k}\,,  \nn\\
  f_{C}(q,u) &\equiv \sum_{\tC\in [\cB(C)]} q^{|\tC|}u^{b(\tC)} \,,
\end{split}
\end{equation}
for a representative of primitive cycles $C$. 

We first note that, 
if we label the elements of $[\cB(C)]$, that is, the set of cycles that reduce to $C$, as $C_j \in [\cB(C)]$ ($j\in \N)$,  
${}^\forall \tC\in [\cB(C^d)]$ can be uniquely written as 
\[
\tC = C'_{j_1}\cdots C'_{j_d}\,,
\]
where each $C'_{j_i}$ is an equivalent cycle to $C_{j_i}$. 
This means that, 
as far as we consider such quantity that is irrelevant to the cyclic rotation of the cycles in $[\cB(C)]$, 
we can identify an element of $[\cB(C^d)]$ as a word of length $d$ generated by the elements of $[\cB(C)]$. 
Therefore, we rather consider words generated by $[\cB(C)]$, instead of cycles, in the following. 

We then consider the summation of all possible words with length $n$ made of the elements of $[\cB(C)]$,
which is given by $\left(\sum_{\tC\in[\cB(C)]} \tC\right)^n$. 
Since any word can be expressed as a positive power of a primitive word in general, 
we can describe this quantity as 
\begin{equation}
\left(\sum_{\tC\in[\cB(C)]} \tC\right)^n = \sum_{k|n} \sum_{\tC\in P_k}\tC^{\, n/k}\,,
\label{eq:tochu2}
\end{equation}
where $P_k$ denotes the set of primitive words of length $k$. 
Since an element of $[\cB(C^k)]$ is originally a set of representatives of primitive cycles,
it is also primitive in the sense of words. 
Therefore 
there are $d$ different words which are equivalent to $\tC\in [\cB(C^k)]$. 
If we define a mapping $t(C)$ by 
\[
t(C)=q^{|C|}u^{b(C)}\,,
\]
it is obvious that $t(C)=t(C')$ if $C\sim C'$.
So we do not need to care about the differences between words and cycles.
Therefore, by acting this mapping to both side of \eqref{eq:tochu2}, we finally obtain 
\begin{equation}
\left(\sum_{\tC\in [\cB(C)]} q^{|\tC|}u^{b(\tC)}\right)^n 
= \sum_{k|n} k \sum_{\tC\in [\cB(C^k)]}(q^{|\tC|}u^{b(\tC)})^{n/k}\,, \nn
\end{equation}
which is nothing but the equality \eqref{eq:resultfC}.

\bibliographystyle{unsrt}
\bibliography{refs}

\end{document}